\documentclass[modern]{aastex7}
\usepackage[utf8]{inputenc}
\usepackage{amsmath}
\usepackage{amssymb}
\usepackage{enumitem}
\usepackage{mathtools}
\usepackage{tipx}

\usepackage{hyperref}

\newcommand{\holder}{{H{\"o}lder }}

\begin{document}

\title{X-ray Spectroscopy via Temporal Decomposition}

\author[0000-0003-2165-8314]{William Setterberg}
\affiliation{University of Minnesota}
\email{sette095@umn.edu}
\author[0000-0001-7092-2703]{Lindsay Glesener}
\affiliation{University of Minnesota}
\email{glesener@umn.edu}

\begin{abstract}
We present a novel way to fit solar flare X-ray spectra
    that offers more sensitivity to physical flare parameters
    than traditional approaches to spectroscopy.
We decouple physically distinct emission types in
    solar flare X-ray spectra using timing behaviors,
    a technique we call time-decomposed spectroscopy.
By fitting the shapes of particular time series
    to others across a time interval,
    we extract X-ray emission of distinct physical origins before any
    forward modeling.
We perform spectroscopy on the original and time-decomposed spectra
    and find good agreement,
    with physical reasons for the few disagreements.
In general,
    the time decomposition technique provides more precise results
    than those of traditional spectroscopy.
The thermal and nonthermal energies are better constrained by
    more than an order of magnitude using the time decomposition approach,
    relative to traditional spectroscopy.
We explain mathematically how and why the technique works
    using the multifractal formalism,
    and show that fractality is one way to
    choose appropriate light curves for this analysis.
We speculate about applications to different wavelengths
    for solar data analysis,
    as well as applications to other physics subfields and domains.

\end{abstract}

\section{Introduction}

Solar flares energize particles to keV energies via heating and nonthermal mechanisms.
Particles accelerated to energies much greater than the thermal energy
    (a.k.a. nonthermal particles) make up a significant fraction of the flare energy budget
    \citep{lin-hudson-energy}.
Generally,
    nonthermal particles in solar flares follow a power law distribution in energy,
    so most of their energy is allocated to the lower-energy nonthermal particles.
However,
    measuring the low-energy behavior of these nonthermal electron populations is difficult,
    and remains the largest contributor to the nonthermal energy uncertainty \citep{pascal-energies}.
The transition from hot plasma X-ray emission (thermal bremsstrahlung)
    to e.g. cold, thick target power-law nonthermal emission \citep{brown-cold-thick}
    occurs in an energy range which is dominated by thermal bremsstrahlung emission.
In other words,
    it is hard to fit the low-energy cutoff parameter $E_c$ assumed by
    the cold,
    thick target model using traditional spectroscopy.
Current spectroscopy methods lead to degenerate solutions even with the simplest composite models,
    and without better tools to explore X-ray data,
    we cannot confidently investigate more complex physics,
    such as the low-energy shape of nonthermal particle populations.
Better understanding the low-energy behavior of the electron population
    would yield physical insights for observers,
    modelers,
    and experimentalists alike about solar flare physics.

Motivated by this,
    we present a new approach to spectroscopy.
We exploit timing differences in thermal and nonthermal X-ray time profiles to aid in spectroscopy.
Often, nonthermal X-ray fluctuations occur
    on timescales of a few seconds \citep{grigis-benz},
    and on shorter than one second
    in select cases
    \citep{kiplinger-spikes, qiu-case, qiu-statistical, trevor-gbm}.
The nonthermal timing fluctuations are relatively fast
    because they are directly associated with electron acceleration and transport.
In contrast,
    variations in thermal bremsstrahlung emission occur on timescales of
    minutes to tens of minutes,
    and sometimes appear as quasi-periodic pulsations \citep{afino-first, laura-qpp}.
These timescales are longer because the plasma takes time to heat.


We use the differences in these timescales to sift emission per energy bin
    by temporal behavior.
We describe this approach in precise equations,
    which are straightforward to apply.
We apply this approach to two flares,
    Flare F1 observed by the Spectrometer/Telescope for Imaging X-rays
    (STIX, \cite{stix-instr}),
    and Flare F2 observed by the Reuven Ramaty High-Energy Solar Spectroscopic Imager
    (RHESSI, \cite{rhessi-instr}).
RHESSI uses time modulation to perform Fourier imaging,
    while STIX uses spatial modulation.
In both cases,
    the decomposition produces physically reasonable results.

We compare energy computation results and show that the time-decomposition
    technique places more precise bounds on the nonthermal energy deposited,
    and slightly more precise bounds on the flare plasma thermal energy.
We discuss the physics of how timing can lift some degeneracies
    intrinsic to traditional spectroscopy.
We discuss potential applications to multiple wavelengths
    and the irrelevance of periodic behavior.
Finally,
    we explain how the multifractal formalism may be used
    in tandem with the time decomposition technique.

\section{Temporal decomposition of X-ray time series}
\label{sec:tempspec}
\begin{figure}
    \centering
        \includegraphics[width=\textwidth]{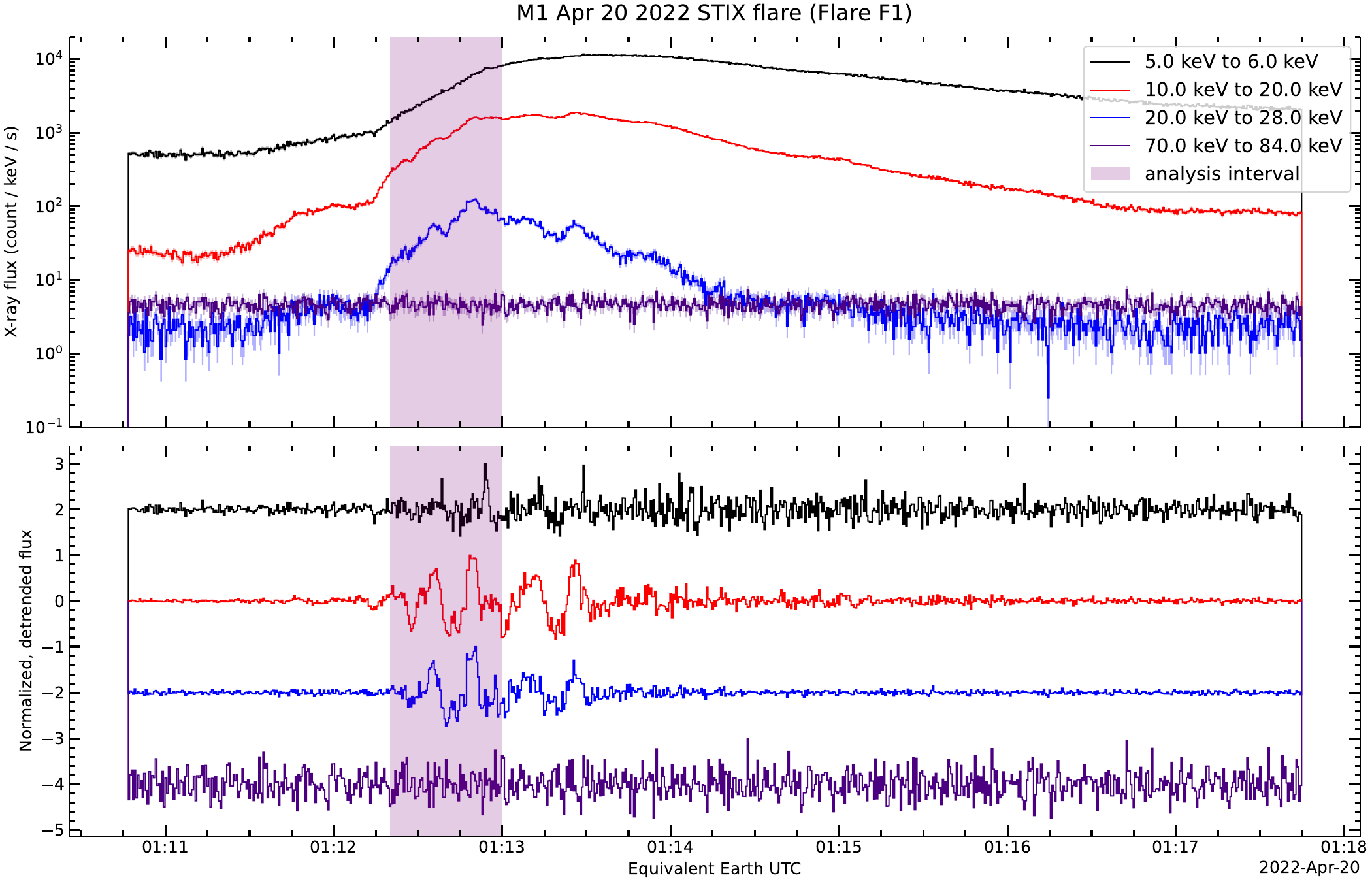}
    \caption{
        X-ray light curves for an M-class flare observed by Solar Orbiter/STIX;
        we call this \textbf{Flare F1}.
        The pink region is the analysis interval.
        Times are adjusted for 1 AU photon arrival time,
            so they are a few minutes later than the UTC times recorded aboard Solar Orbiter.
        The red and blue curves correspond to higher-energy X-rays,
            and the detrended plots show how these exhibit relatively
            more intense variations within the analysis interval than
            the lower energy (black) light curve does.
        The indigo curve is detector background,
            entirely comprised of uncorrelated shot noise.
        The detrended curves are high-pass filtered.
        The detrending was performed using a zero-lag fifth-order
            low-pass Butterworth filter with a cutoff frequency corresponding
            to 20s \citep{butterworth, gustafsson-filter-boundaries}.
    }
    \label{fig:stix-lc}
\end{figure}

\subsection{An intuitive description}

The temporal decomposition method uses differences in timing characteristics between
    solar thermal and nonthermal X-ray bremsstrahlung
    to decompose spectra along the energy dimension.
First,
    we illustrate the basic approach by means of example.
\autoref{fig:stix-lc} shows X-ray light curves for Flare F1,
    observed on 2022 April 20 by STIX.
The black light curve in the figure is dominated by emission from thermal plasma,
    while the blue light curve is dominated by nonthermal bremsstrahlung,
    which exhibits more abrupt changes associated with multiple energy release
    events during the flare.
During the impulsive phase,
    the red curve is a mix of thermal and nonthermal bremsstrahlung,
    and exhibits some short- and some longer-timescale behavior.
 Traditionally,
    the emission over an analysis time interval would be integrated together
    and fit simultaneously with a set of spectral models to determine how much thermal
    and non-thermal emission is present.
Here, we present a different approach.

\begin{figure}
    \centering
        \includegraphics[width=\textwidth]{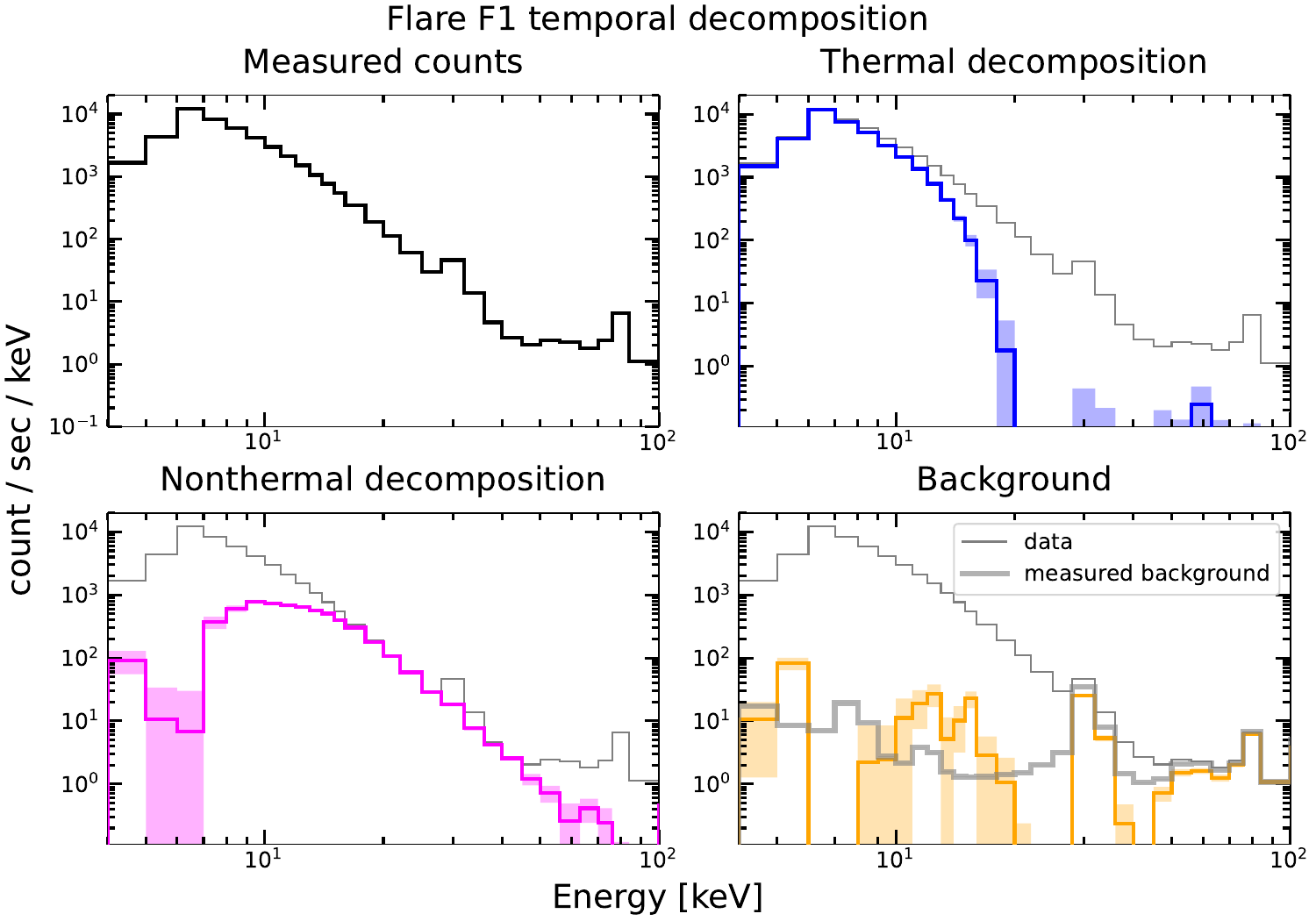}
        \caption{
            Time-decomposed spectra from Flare F1.
                The first panel shows the measured data;
                the second shows the time-decomposed thermal emission (blue);
                the third shows the time-decomposed nonthermal emission (magenta);
                the fourth panel shows the measured (gray)
                and time-decomposed (orange) backgrounds.
            Using only temporal variation to define the pseudobases in \autoref{eq:tedec},
            the spectra capture the background,
                thermal emission,
                and nonthermal emission with physically realistic spectral shapes,
                despite no spectroscopy yet being performed.
            In the fourth panel,
                a measured background from 19 April 2022 is overplotted
                on the decomposition.
            The decomposition technique captures the background well
                when its intensity is at least the order of that of the flare.
        }
        \label{fig:compare-decompositions}
\end{figure}

The key idea is to combine
    light curves across time and
    apply those combination coefficients across energy.
In the case of \autoref{fig:stix-lc}:
    we combine the thermal and nonthermal (black and blue)
    light curves across time with some weights
    to reconstruct the shapes of other light curves (like the red one).
The weights of the blue and black curves, per energy, yield projection coefficients.
Then,
    we sum the light curves across the analysis time interval into a counts vs. energy spectrum,
    and multiply the resulting energy spectra by those same time projection coefficients;
    this leverages temporal behavior to separate energetic components of the emitting plasma populations.
\textbf{The result is that we separate the thermal and non-thermal emission before doing any spectral fitting;
    an example is shown in \autoref{fig:compare-decompositions}.}

\subsection{The decomposition formalism}
We outline the formalism as follows.
The relevant code is freely available online\footnote{
    \href{https://github.com/settwi/tedec}{GitHub: decomposition package}; for paper-specific analysis questions, please contact the authors.
}.
First, we choose at least two pseudobasis vectors which
    are used to decompose the data across time\textemdash
    e.g. the black and blue light curves in \autoref{fig:stix-lc}.
The pseudobases should be as representative as possible of the different
    populations present in the data.
In this case,
    we choose light curves we believe to be dominated by thermal and nonthermal emission,
    respectively.
We call these light curves time pseudobases;
the thermal pseudobasis is $T$ and the nonthermal one is $N$.

Next,
    we normalize the light curves across time to address
    the intensity difference between lower- and higher-energy X-ray emission,
    which can be several orders of magnitude due to the steepness of solar flare X-ray spectra.
We combine light curves along the energy axis to form an 
    $(E, t)$ spectrogram matrix with elements $S_{ij}$.
The first index corresponds to energy and the second to time.
In the case of STIX,
    the index $i \in \{0, 1, \dots, 29\}$
    corresponds to its science energy bins,
    and $j \in \{0, 1, \dots, M\}$ the time bins across
    the purple interval in \autoref{fig:stix-lc}.
For unnormalized data we use upper case letters, like $S$, $N$, $T$.
For normalized spectrogram elements,
    we use lower-case letters,
    as in \autoref{eq:tedec-norm},
    where $s_{ij}$ is a generic spectrogram element which has been normalized by
    total counts across a fixed time interval.
\begin{equation}
    s_{ij} = \frac{S_{ij}}{\sum_k S_{ik}}
    \label{eq:tedec-norm}
\end{equation}
The values $s_{ij}$ are unitless,
    so any type of data (e.g. flux, counts, count rate) 
    may be used with this technique.

Next, we decompose all normalized light curves $s_{ij}$ across time.
The normalized nonthermal pseudobasis is $n_{j}$ and thermal pseudobasis is $t_{j}$.
The pseudobases do not have energy indices because they are implicitly associated with energy bins.
We perform a multilinear regression across time as in \autoref{eq:tedec}.
The constant offset term $1_j$ is normalized in the same way as $N$ and $T$ and written out explicitly
    to avoid confusion.
The coefficient $\alpha_i$ is the portion of the arbitrary spectrogram element $s_{ij}$ which behaves
    most like the thermal pseudobasis,
    $\beta_i$ is that which is most like the nonthermal pseudobasis,
    and $\gamma_i$ is a constant offset term which can account for the background
    if it is relatively constant across time.
This technique can also account for time-varying backgrounds,
    which we describe later.
\begin{equation}
    s_{ij} \approx \alpha_i t_j + \beta_i n_j + \gamma_i \frac{1_j}{\sum_k 1_k}
    \label{eq:tedec}
\end{equation}

We estimate the pseudobasis vectors from the data using low- and high-energy 
    light curves.
Model predictions or external data sources may also form the pseudobases.
If the element $S_{ij}$ is not consistent to within a few percent,
    the technique is being applied inappropriately,
    and the pseudobases need to be reexamined.

After decomposing the spectra,
    we take the energy-dependent coefficients $\alpha_i$ $\beta_i$ $\gamma_i$,
    and use them to project the temporal behavior of the time series across the energy axis
    of the spectrogram, as in \autoref{eq:tedec-reproject}.
\begin{equation}
    S_{ij} \approx 
        {\color{blue}       \left(\alpha_i \sum_k S_{ik}\right)  }  \frac{T_j}{\sum_k T_k} +
        {\color{magenta}    \left(\beta_i \sum_k S_{ik}\right)  }  \frac{N_j}{\sum_k N_k} +
        {\color{orange}     \left(\gamma_i \sum_k S_{ik}\right)}  \frac{1_j}{\sum_k 1_k} 
    \label{eq:tedec-reproject}
\end{equation}
\textbf{The colored/parenthesized portions of \autoref{eq:tedec-reproject}
        are the time-decomposed energy spectra.}
Each normalized spectrogram element is explicitly written for notational consistency.
The summation in the denominator of \autoref{eq:tedec-norm} has been multiplied to the right
    to form time-independent projections across energy of the various pseudobases.

We plot the colored coefficients from \autoref{eq:tedec-reproject} as a function
    of energy for Flare F1 in \autoref{fig:compare-decompositions}.
For all of the data in this work,
    we assume Gaussian statistics.
The errors are propagated via a Monte Carlo technique,
    wherein the original light curves are sampled using the count error distribution,
    and the resampled data is used to recompute the result many times.
In \autoref{fig:compare-decompositions},
    we observe that the blue and magenta curves have spectral shapes which
    closely match typical thermal and nonthermal emission spectra,
    and find that the background decomposition captures the ``true''
    background when the emission is not source-dominated;
    the colors of the coefficients in \autoref{eq:tedec-reproject}
    correspond to the colors of the plots.
In \autoref{sec:results},
    we show that the decomposed spectra represent physically meaningful
    components of the flare X-ray emission.

More generally,
    one may use any number of pseudobases in \autoref{eq:tedec} and \autoref{eq:tedec-reproject}.
In the case of RHESSI data,
    this is useful for fitting the background,
    as it varies with time.
A high-energy light curve can be used as the background pseudobasis across time
    instead of a constant offset.
A constant like the orange term in \autoref{eq:tedec-reproject} may be added
    if desired or required,
    however a pseudobasis can often play the role of the background.

Furthermore,
    one may sum together one or more light curves to form a new pseudobasis,
    so long as the normalizations in \autoref{eq:tedec-reproject} are respected.
For the solar X-ray case,
    one may wish to sum  a few thermal bremsstrahlung-dominated light curves
    and a few nonthermal bremsstrahlung-dominated light curves together to form
    energy-averaged thermal and nonthermal pseudobases.

After decomposing the spectra using timing information,
    we perform spectroscopy on the independent components.
The temporal decomposition allows us to fit the thermal and nonthermal spectra separately,
    letting the models vary independently from one another,
    which reduces the degeneracy of spectral fits.

\section{Results}
\label{sec:results}

We apply time-decomposed spectroscopy to two GOES M-class solar flares.
Flare F1\textemdash which has been examined some in prior sections\textemdash
    was observed by STIX on 20 April 2022 (M1 GOES class),
    and Flare F2 by RHESSI on 30 July 2011 (M9 GOES class).
Both flares exhibit nonthermal emission in the $> \sim 25$ keV energies.

\begin{table}
\centering
\begin{tabular}{c|c|c|c|c}
    Flare & Technique & Parameter & Best fit value & 95\% posterior interval \tabularnewline \hline
    F1 & traditional & $T$ & 23 MK & (16, 25) MK \tabularnewline
    F1 & traditional & $EM$ & $54 \times 10^{46}$ cm\textsuperscript{-3} &
        $(46, 160) \times 10^{46}$ cm\textsuperscript{-3} \tabularnewline
    F1 & traditional & $\varphi_e$ & $3.0 \times 10^{35}$ electron s\textsuperscript{-1} &
        $(1.9, 35) \times 10^{35}$ electron s\textsuperscript{-1}\tabularnewline
    F1 & traditional & $\delta$ & 6.5 & (6.3, 6.8) \tabularnewline
    F1 & traditional & $E_c$ & 21 keV & (14, 23) keV \tabularnewline
    \hline
    F1 & decomposition & $T$ & 20 MK & (20, 21) MK \tabularnewline
    F1 & decomposition & $EM$ & $88 \times 10^{46}$ cm\textsuperscript{-3} &
        $(77, 100) \times 10^{46}$ cm\textsuperscript{-3}\tabularnewline
    F1 & decomposition & $\varphi_e$ & $3.1 \times 10^{35}$ electron s\textsuperscript{-1} &
        $(2.8, 3.5) \times 10^{35}$ electron s\textsuperscript{-1}\tabularnewline
    F1 & decomposition & $\delta$ & 6.3 & (6.1, 6.5) \tabularnewline
    F1 & decomposition & $E_c$ & 22 keV & (21, 23) keV \tabularnewline
    \hline\hline
    F2 & traditional & $T$ & 22 MK & (21, 23) MK \tabularnewline
    F2 & traditional & $EM$ & $2.1 \times 10^{49}$ cm\textsuperscript{-3} &
        $(1.6, 2.9) \times 10^{49}$ cm\textsuperscript{-3}\tabularnewline
    F2 & traditional & $\varphi_e$ & $5.7 \times 10^{35}$ electron s\textsuperscript{-1} &
        $(2.6, 390) \times 10^{35}$ electron s\textsuperscript{-1}\tabularnewline
    F2 & traditional & $\delta$ & 6.1 & (5.9, 6.4) \tabularnewline
    F2 & traditional & $E_c$ & 23 keV & (10, 27) keV \tabularnewline
    \hline
    F2 & decomposition & $T$ & 22 MK & (21, 23) MK \tabularnewline
    F2 & decomposition & $EM$ & $3.2 \times 10^{49}$ cm\textsuperscript{-3}  &
        $(2.6, 3.9) \times 10^{49}$ cm\textsuperscript{-3}\tabularnewline
    F2 & decomposition & $\varphi_e$ & $1.2 \times 10^{35}$ electron s\textsuperscript{-1} &
        $(0.97, 1.7) \times 10^{35}$ electron s\textsuperscript{-1}\tabularnewline
    F2 & decomposition & $\delta$ & 5.2 & (5, 5.3) \tabularnewline
    F2 & decomposition & $E_c$ & 29 keV & (26, 32) keV \tabularnewline
\end{tabular}
    \caption{
        The best-fit (minimized $\chi^2$) parameters for flares F1 and F2,
            for both the traditional and time-decomposed fits.
        Errors and Markov chain Monte Carlo (MCMC)
        medians are also reported in the corner plots in \autoref{sec:appendix-corners}.
    }

    \label{tab:params}
\end{table}

\subsection{Flare F1: 20 April 2022 flare observed by SolO/STIX}
First,
    we perform traditional and time-decomposed spectroscopy on Flare F1,
    from \texttt{2022-04-01T01:12:20Z} to \texttt{01:13:00Z}.
The flare X-ray emission (\autoref{fig:stix-lc}) is decomposed using the technique described in
    \autoref{sec:tempspec}.
The thermal pseudobasis is a sum of the three energy bins centered around 6 keV
    which capture the emission lines of highly-ionized iron present in the solar corona.
The nonthermal one is a sum of three energy bins centered around 19 keV,
    where the nonthermal emission is dominant.

The counts and count errors are livetime-corrected,
    and errors are assumed to be normally distributed in all cases.
We account for error on the counts due to Poisson statistics,
    as well as counts and livetime due to X-ray count and trigger count compression,
    which STIX employs due to its small downlink volume \citep{stix-instr}.

Before temporal decomposition, no systematic error is applied to the data.
For both traditional and time decomposed spectroscopy,
    we apply a 20\% systematic error to the smallest energy bin,
    and a 10\% systematic error to all other energy bins.
The systematic error is added in quadrature:
    $\sigma_\text{tot}^2 = \sigma_\text{data}^2 + (\sigma_\text{sys}\cdot\text{counts})^2$.
The larger systematic uncertainty on the lowest energy bin reflects
    the relative complexity of characterizing the low-energy
    behavior of X-ray detectors\textemdash
    see, e.g., \cite{ishikawa-foxsi3}\textemdash
    as well as the evolving understanding of the STIX low-energy response.

\begin{figure}
    \centering
    \includegraphics[width=0.95\textwidth]{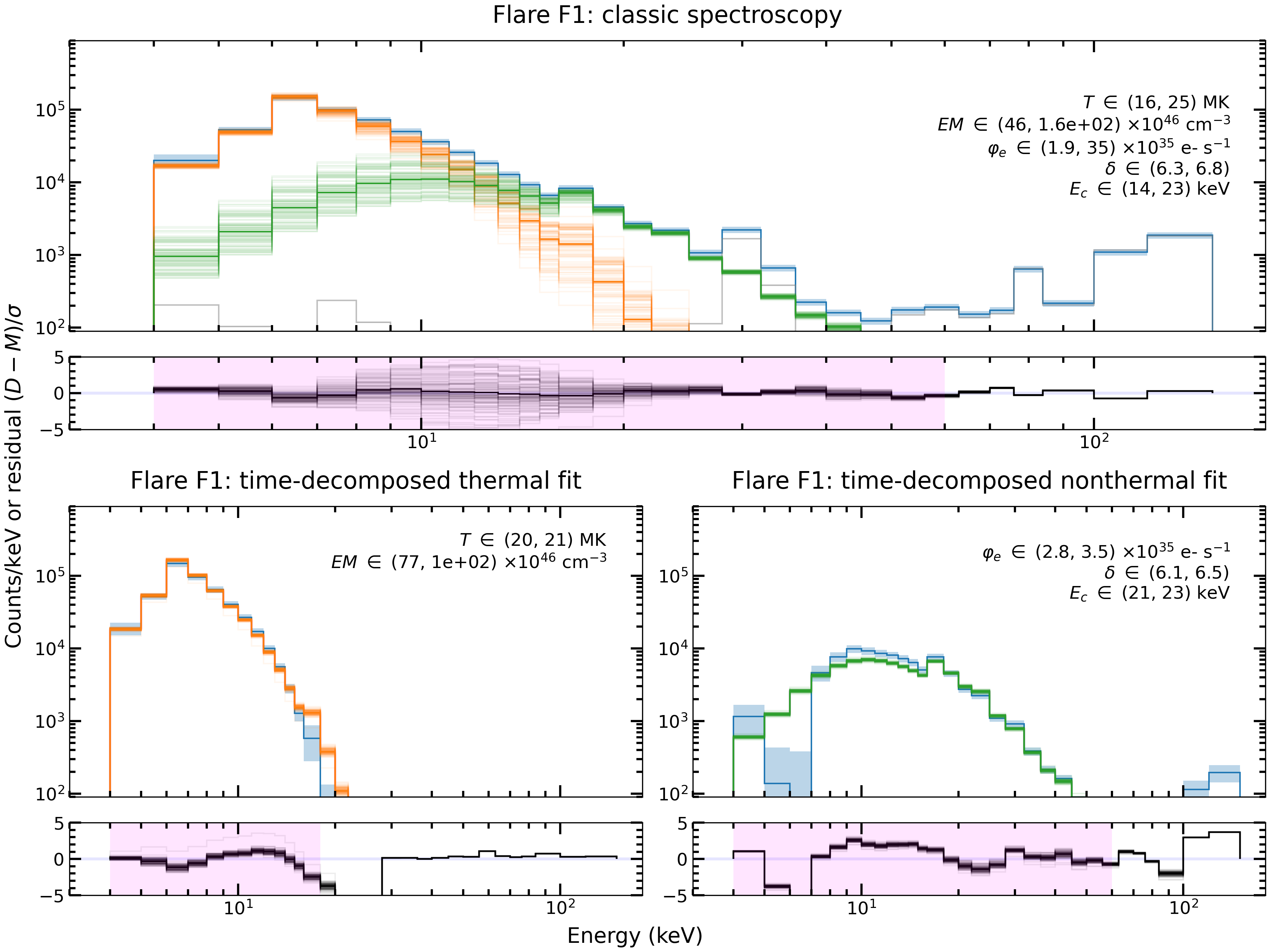} 
    \caption{
        Comparison of traditional vs. time-decomposed spectroscopy for Flare F1 (M1 class 20 Apr 2022).
        The blue lines are the data,
            orange lines are the thermal bremsstrahlung fits,
            and green lines are the cold thick target fits.
        The pink regions in the residual plots indicate the energy fitting ranges.
        The top panel shows traditional spectroscopy and the bottom two panels
            show the results from the time-decomposed fits.
        The plots are annotated with 95\% $\chi^2$ parameter intervals.
        The spurious counts in the thermal decomposition at high energies are nonphysical
            and not included in the fit range.
    }
    \label{fig:stix-decomp-vs-traditional}
\end{figure}

After decomposition,
    we fit the blue curve (thermal decomposition)
    and magenta curve (nonthermal decomposition) in
    \autoref{fig:compare-decompositions} completely independently.
For the background pseudobasis, we sum three energy bins closest to 80 keV.
We compare the background decomposition to the pre-flare background,
    and find that the background spectrum is well-modeled by the time decomposition
    when it is at least as intense as the flare emission.

We fit the livetime-adjusted counts using $\chi^2$ minimization and run MCMC using the
    \texttt{emcee} Python package \citep{emcee} on the $\chi^2$ value
    to obtain parameter uncertainties;
    corner plots of the parameter posterior distributions are in \autoref{sec:appendix-corners}.
We also fit the X-ray spectra using typical X-ray spectroscopy where both the thermal
    and nonthermal components are fit simultaneously, for comparison.
The results for Flare F1 are displayed in \autoref{fig:stix-decomp-vs-traditional}.

The model parameter intervals annotated on \autoref{fig:stix-decomp-vs-traditional}
    are the 95\% $\chi^2$ intervals for each spectroscopy method.
All intervals overlap at the 95\% level.
Notably, however,
    the time-decomposed intervals are more precise than
    those of traditional spectroscopy.
This better constrains the nonthermal electron power,
    as we discuss in \autoref{sec:discussion}.
The thermal bremsstrahlung function fit begins to diverge around $\sim18$ keV,
    where the nonthermal flux is far brighter than the thermal flux.
In the case of traditional spectroscopy,
    this energy would be totally dominated by nonthermal emission.
We use this physical reasoning to restrict the fit range to $\le 18$ keV,
    and ignore spurious counts above this energy.

\subsection{Flare F2: 30 July 2011 flare observed by RHESSI}
\label{subsec:rhessiflare}
\begin{figure}
    \centering
    \includegraphics[width=0.95\textwidth]{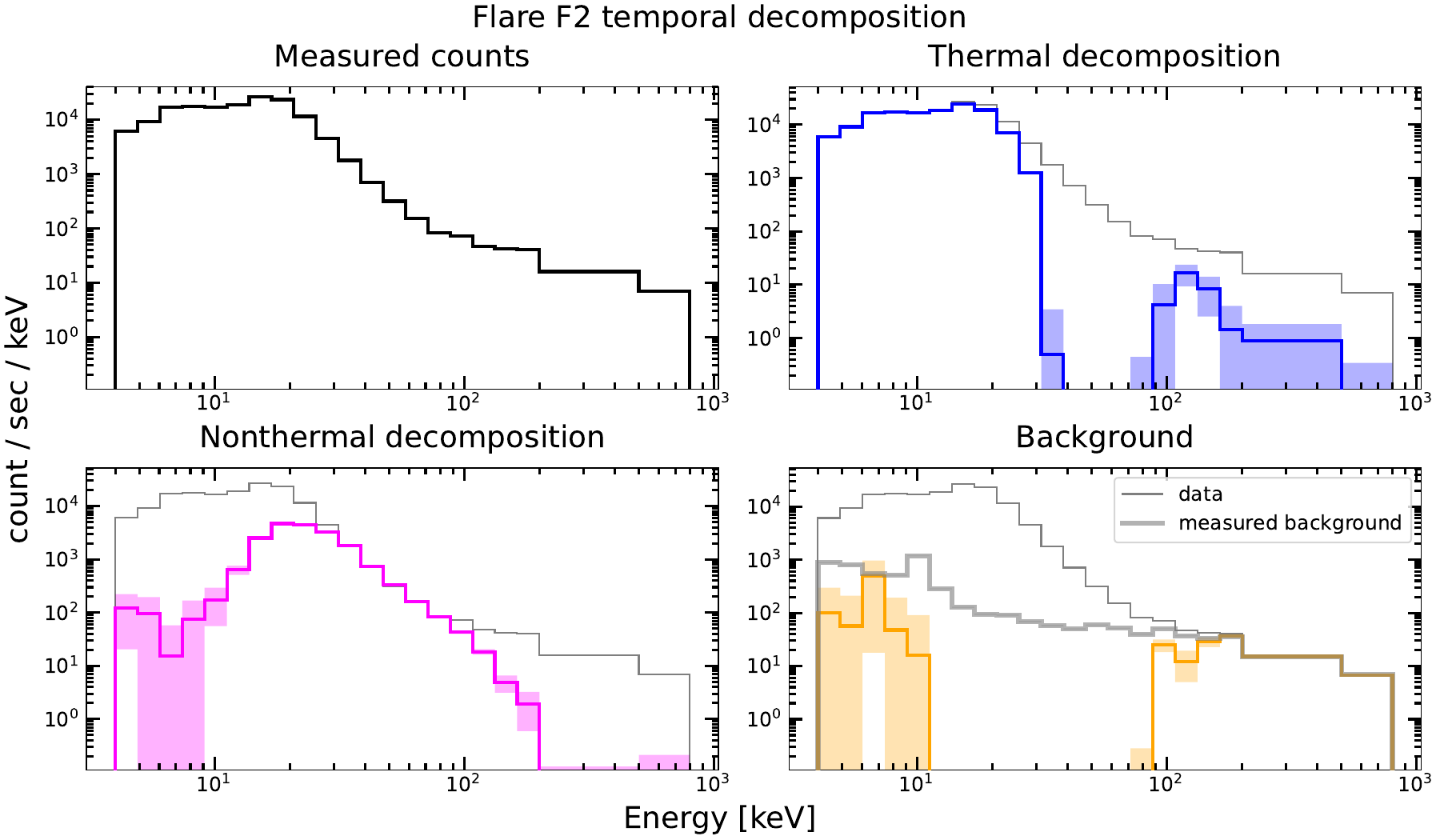} 
    \caption{
        The data and time-decomposed components for Flare F2.
        The first panel shows the measured data;
        the second shows the time-decomposed thermal emission;
        the third shows the time-decomposed nonthermal emission;
        the fourth panel shows the measured and time-decomposed backgrounds.
        In the fourth panel,
            a pre-flare background is overplotted on the decomposition.
        As in the case of Flare F1,
            the decomposition technique captures the background well
            when its intensity is at least the order of that of the flare,
            despite the time-varying nature of RHESSI's background.
    }
    \label{fig:rhessi-decomp-components}
\end{figure}

We apply the same temporal decomposition technique to Flare F2,
    across a two-minute interval near the peak of the flare,
    after both RHESSI's thick and thin attenuators have been inserted,
    from \texttt{2011-07-30T02:08:20Z} to \texttt{2011-07-30T02:10:20Z}.
We choose a sum of three bins near 6 keV as the thermal pseudobasis,
    the same for three bins near 80 keV for the nonthermal pseudobasis,
    and use a large energy band from 200 to 500 keV as the background pseudobasis.
The decomposed spectra are presented in \autoref{fig:rhessi-decomp-components}.

\begin{figure}
    \centering
    \includegraphics[width=0.95\textwidth]{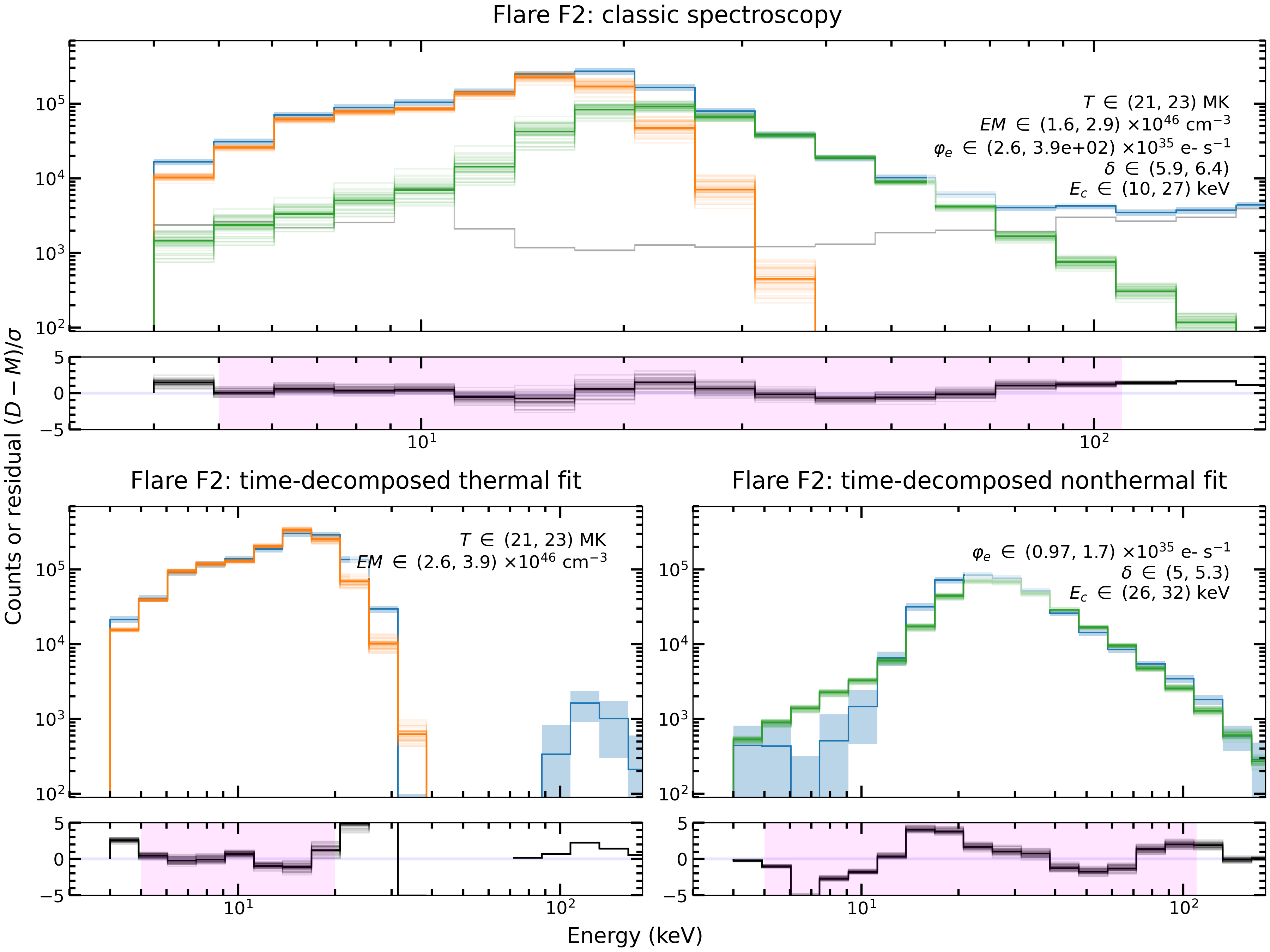} 
    \caption{
        Comparison of traditional vs. time-decomposed spectroscopy for Flare F2 (M9 class 30 Jul 2011).
        The blue lines are the data,
            orange lines are the thermal bremsstrahlung fits,
            and green lines are the cold thick target fits.
        The pink regions in the residual plots indicate the energy fitting ranges.
        The gray line in the first panel is the background.gray line in the first panel is the background.gray line in the first panel is the background.
        The top panel shows traditional spectroscopy and the bottom two panels
            show the results from the time-decomposed fits.
        The plots are annotated with 95\% $\chi^2$ parameter intervals.
        There are many MCMC samples overplotted to give a sense of model variation.
        The spurious counts in the thermal decomposition at high energies are nonphysical
            and not included in the fit range;
            the corresponding high-energy residuals are small because of the very large
            error on these nonphysical counts.
    }
    \label{fig:rhessi-decomp-vs-traditional}
\end{figure}

A light curve of this flare in RHESSI, Fermi/GBM,
    and GOES XRS may be found in \cite{trevor-gbm},
    Figure 1.
Again we fit livetime-adjusted counts and use thermal bremsstrahlung and cold thick target models.
Errors are assumed to be Gaussian and are taken from an \texttt{hsi\_spectrum} IDL object output.
No systematic uncertainty is added before the decomposition.
Counts and errors are livetime corrected (divided by the livetime fraction).
For both traditional and decomposed fitting,
    a uniform 10\% systematic uncertainty is added
    in quadrature to the initial counts errors before spectroscopy.
The temporally-decomposed and traditional spectroscopy results are displayed in \autoref{fig:rhessi-decomp-vs-traditional}.

As is the case with STIX data,
    the temporal decomposition technique produces physically reasonable results.
The systematic variation in the temporally decomposed $\chi^2$ is more notable
    than in the traditional spectroscopy case.
All parameters except the spectral index and electron flux overlap at the 95\% $\chi^2$ level.

\begin{figure}
    \centering
    \includegraphics[width=0.95\textwidth]{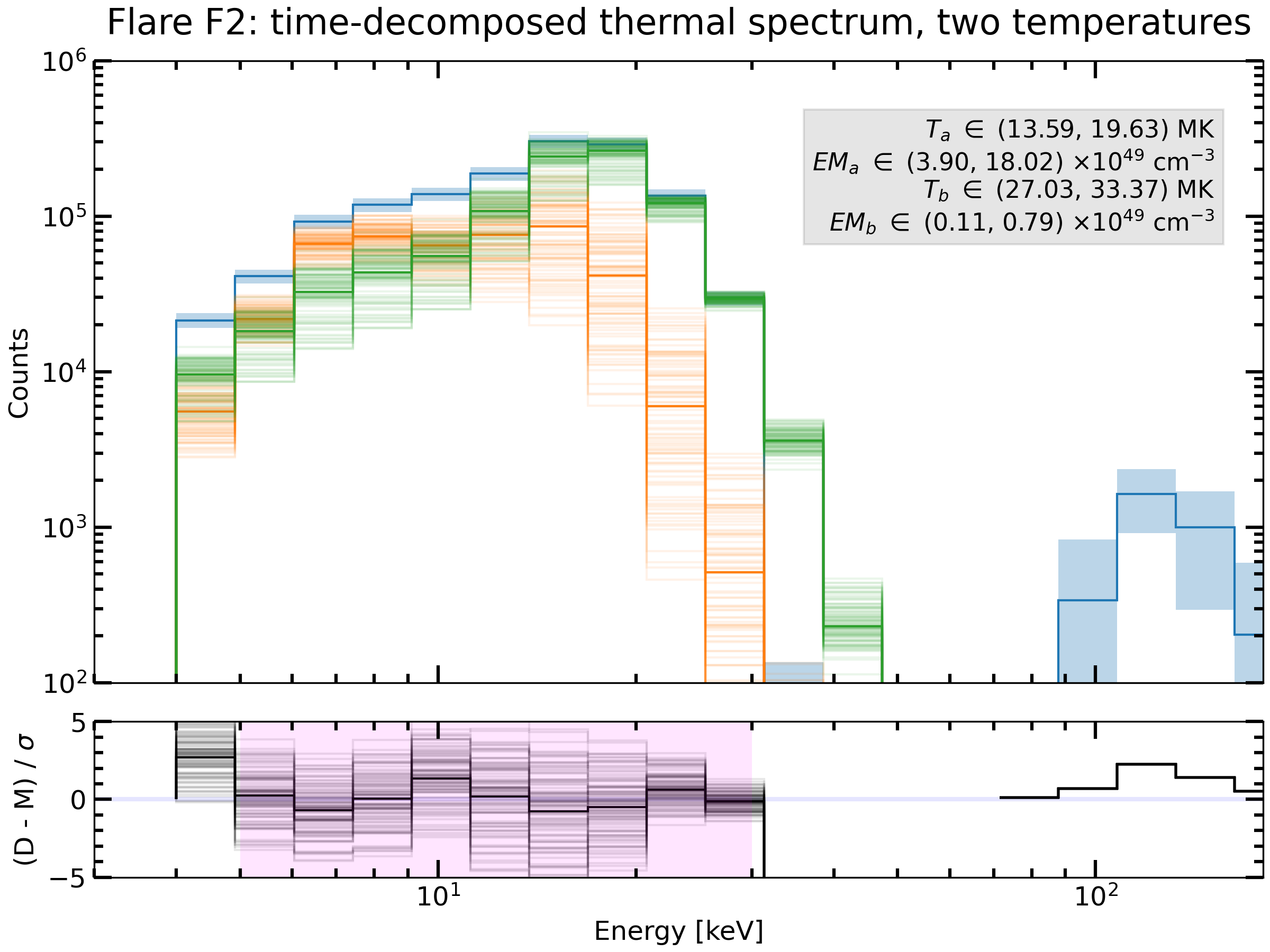} 
    \caption{
        We fit two isothermal bremsstrahlung emission functions to the time-decomposed
            Flare F2 thermal spectrum.
        The time-decomposed data is blue; the high-energy portion is unphysical and ignored.
        We achieve better model/data agreement when fitting
        the two components:
            one is cool and high emission measure (orange),
            and the other is hot and lower emission measure (green).
    }
    \label{fig:rhessi-double-thermal}
\end{figure}

There are a few notable distinctions between the fitting technique residuals.
In the thermal model fits,
    the decomposed model underestimates the data $>\sim 20$ keV.
This suggests a second,
    hotter source of X-rays may be present.
We performed a fit with a two-temperature model
    on the thermal time-decomposed spectrum;
    this assumes the thermal components vary similarly in time.
We found good
    agreement up to 30 keV,
    with one component being about 18 MK and the other 30 MK (\autoref{fig:rhessi-double-thermal}),
    and this greatly improved the fit residuals.
The emission measure of the superhot 30 MK component is
    an order of magnitude smaller than that of the 18 MK component.

Models with more free paramters generally fit data better.
However,
    the hotter temperature we found is in agreement
    with prior studies of so-called superhot plasma,
    which is hypothesized to be the result of a small volume
    of low-density plasma being heated directly at the
    particle acceleration site in the corona \citep{amir-superhot}.
The second temperature we find is in line with the statistics presented in \cite{amir-superhot},
    so this provides evidence that a second temperature component is likely present.

In both spectroscopy approaches,
    there is a deficit in the high-energy ($> \sim 70$ keV)
    nonthermal component.
This suggests a second electron population may be present
    in the form of a broken power law in electron space.
The presence of such a break would be consistent
    with prior statistical studies of high-energy RHESSI flares \citep{meriem-breaks}.
In contrast,
    there is disagreement around the low-energy shape of the nonthermal spectrum,
    as is also the case in Flare F1.
This may be due to the presence of a different low-energy behavior between the two
    electron populations,
    or because the low-energy cutoff is not actually a sharp cutoff.
It should be easier to see the true low-energy nonthermal shape in the time-decomposed case,
    as well as any ensuing disagreement between model and data.
This idea is supported by the better constraint we place on the nonthermal parameters:
    in the traditional case,
    the electron flux spans two orders of magnitude,
    but is much better constrained in the time-decomposed fits.

Furthermore,
    traditional spectroscopy is very sensitive to background subtraction when
    the background intensity is comparable to the source intensity.
The time decomposition technique inherently subtracts the background from the data,
    which provides better sensitivity in these regimes,
    and lets us follow the nonthermal behavior to higher and lower energies more confidently.

\section{Discussion}
\label{sec:discussion}
Here, we discuss the time-decomposition spectroscopy results in more detail.
First,
    we quantify the energy content inferred from both spectroscopy techniques,
    and find they are consistent,
    but that the time-decomposition technique is much more precise.
Next,
    we describe why using timing information lifts spectroscopy degeneracies,
    and go on to discuss intrinsic biases in both the traditional and time-decomposition techniques.
Finally,
    we speculate about applications to cross-instrument analysis
    and entirely different types of data.

\subsection{Energetics}
Here we quantify the energy content of Flares F1 and F2,
    using results from traditional and time-decomposed spectroscopy.
Because the time-decomposed technique leads to smaller
    95\% parameter intervals,
    the bounds on thermal and nonthermal energy are better than in the traditional case\textemdash
    sometimes significantly so.

We compute the volume of Flares F1 and F2
    using geometry to estimate the thermal energy of the emitting plasma.
While both flares were unocculted from the point of view of the respective HXR instrument,
    Flare F1 was partially occulted from the point of view of SDO/AIA.  
Because of this, we use slightly different approaches to compute the flare volumes.
For Flare F1,
    we take the volume $V = h^3$ where $h$ is the loop height.
We use a geometric approach and knowledge of the flare center from the STIX persepective to 
    estimate the height of the flare using SDO/AIA images.
Taking into account a $4^\circ$ occultation and the visible height in
    an SDO/AIA image,
    we estimate a height $h \approx 13$ Mm,
    and a flare volume $V \approx 2200$ Mm\textsuperscript{3} $ = 2.2 \times 10^{27}$ cm\textsuperscript{3}.
We estimate the thermal energy as $E_\text{th} \approx 2 \cdot \frac32 \sqrt{\text{EM} \cdot fV} k_B T$,
    where $k_B$ is the Boltzmann constant,
    $T$ is the isothermal plasma temperature,
    EM is the emission measure,
    and $f = 1$ is the geometric fill factor.

Flare F2 (M9 GOES class) was located on-disk from the Earth perspective.
We approximate the most prominent flare loop as a semicircle and,
    using data from an AIA 94 \AA$\ $image near the extreme ultraviolet peak,
    estimate the volume of the loop as a bent cylinder,
    finding $V \approx 18000$ Mm\textsuperscript{3} = $1.8 \times 10^{28}$ cm\textsuperscript{3}.

Next we compute the nonthermal power and its time integral.
Both flares appeared on-disk from the X-ray instrument perspectives,
    so we assume a cold, thick target model in both cases.
Our assumed nonthermal electron distribution function is
    $f(E) = f_0 \left(E/E_0\right)^{-\delta}$,
    defined for $E \in [E_\text{lo}, E_\text{hi}]$,
    where $E_0 = 1$ keV is the nondimensionalization energy,
    $f_0$ is the normalization constant,
    and $\delta$ is the electron power law index.
If we take the high-energy cutoff $E_\text{hi} \to \infty$,
    the deposited nonthermal energy across analysis interval $\Delta t$ is
    $E_\text{nth} =
     \frac{1 - \delta}{2 - \delta} \varphi_e E_c \cdot \Delta t$,
    where $\varphi_e$ is the electron flux and $E_c$ the low-energy cutoff
    from spectroscopy.

We compute the thermal and nonthermal energies for both spectroscopy techniques
    for the Flares F1 and F2,
    and tabulate the results in \autoref{tab:energy}.
The 95\% intervals in \autoref{tab:energy} for the decomposition technique are smaller
    than the traditional intervals in all cases.
\textbf{
    Our technique better constrains the Flare F1 nonthermal energy by an order of magnitude,
    and the Flare F2 nonthermal energy by two orders of magnitude.
}

The most notable difference between techniques is the RHESSI nonthermal energy deposited.
The spectroscopy interval occurs while both RHESSI attenuators are inserted.
The attenuators block low-energy photons from both thermal and nonthermal distributions,
    which makes it even more difficult to constrain the low-energy nonthermal behavior.
The decomposition technique may not suffer from this limitation as,
    even with reduced statistics,
    the nonthermal X-ray timing remains distinct from the thermal X-ray timing.
However,
    as seen in \autoref{fig:rhessi-decomp-vs-traditional},
    the nonthermal decomposition has an interesting morphology around 10\textendash20 keV,
    perhaps because the low-energy behavior of the nonthermal electron distribution is poorly modeled.

\begin{table}
\begin{center}
    \begin{tabular}{c|c|c|c}
        Flare + emission & Which technique & Median energy ($10^{30}$ erg) & 95\% $\chi^2$ interval ($10^{30}$ erg) \tabularnewline
        \hline
        F1 thermal & traditional &
            0.50 & 
            [0.31, 0.88] \tabularnewline
        F1 thermal & decomposition &
            0.52 &
            [0.48, 0.58] \tabularnewline
        \hline
        F1 nonthermal & traditional &
            1.0 &
            [0.35, 3.8] \tabularnewline
        F1 nonthermal & decomposition &
            0.54 &
            [0.49, 0.59] \tabularnewline

        \hline\hline

        F2 thermal & traditional &
            5.7 & 
            [4.6, 7.0] \tabularnewline
        F2 thermal & decomposition &
            7.0 &
            [6.0, 8.0] \tabularnewline
        \hline
        F2 nonthermal & traditional &
            7.3 &
            [1.7, 97] \tabularnewline
        F2 nonthermal & decomposition &
            0.94 &
            [0.77, 1.2] \tabularnewline
    \end{tabular}
    \caption{
        The thermal energy is the average plasma energy estimate,
            while the nonthermal energy is the amount of energy deposited assuming a thick target scenario.
        In this table,
            we present the median energy value computed from 6000 MCMC samples
            for two reasons.
        First:
            the traditional spectroscopy methods have parameter posterior distributions
            with large skew.
        Second: the time-decomposition values are unimodal,
            so the difference between the best-fit and median values is negligible.
    }
    \label{tab:energy} 
\end{center}
\end{table}

\subsection{Timing information lifts degeneracies}

Correlations between forward-fitted model parameters
    make it difficult to constrain physical properties
    of the emitting plasma.
The time-decomposed spectroscopy technique lifts some of these degeneracies
    and gives better estimates of physical parameters of the electrons
    participating in the solar flare.

Isothermal bremsstrahlung and cold thick-target bremsstrahlung exhibit
    several correlations, exhibited in \autoref{fig:corner-rhessi-full} (\autoref{sec:appendix-corners}).
In the case of isothermal bremsstrahlung emission,
    the temperature is strongly negatively correlated with the emission measure,
    a common occurrence in solar flare spectroscopy.
There is competition between overall ``brightness'' of the flare (emission measure),
    along with the energetic spread of the assumed Maxwellian electron population
    which is emitting X-rays (temperature).
In the case of the cold target nonthermal bremsstrahlung,
    the cutoff energy $E_c$ and the electron flux $\varphi_e$
    are also strongly negatively correlated:
    both contribute to the total number of electrons which hit the cold target.

When both models are fit simultaneously, other correlations emerge.
By inspection of \autoref{fig:corner-rhessi-full},
    we see that the electron spectral index $\delta$ is weakly correlated with temperature
    and emission measure.
Furthermore,
    the cutoff energy and electron flux are not well-constrained.
The same correlations and poor constraints are present in the corner plots for Flare F1,
    \autoref{fig:corner-stix-full}.

Contrast the simultaneous (thermal + nonthermal) fit results with time-decomposed spectroscopy,
    illustrated in \autoref{fig:corner-rhessi-thermal} and \autoref{fig:corner-rhessi-nonthermal}.
The shapes of the posterior distributions of paremeters
    in the time decomposed case are unimodal and more normally distributed.
The discrepancies between the data and the model found using the time-decomposition method
    indicate the places where our traditional physical models fail.
As described in \autoref{subsec:rhessiflare},
    performing a two-temperature fit on the decomposed data does lead to physically meaningful results,
    like the possible presence of superhot plasma.
The enhancement in the nonthermal decomposition flux relative to the fitted model at low energies may also indicate
    the presence of additional structure in the nonthermal electron population.

\subsection{Intrinsic biases and limitations}

In applying this time-decomposed spectroscopy method,
    we make several simplifying assumptions\textemdash for example,
    that the plasma is isothermal,
    that the cutoff energy is sharp,
    and that there is only one power law of non thermal electrons with a single spectral index.
However,
    all of these assumptions are also present in the traditional method of flare X-ray spectroscopy.
The difference is that our method has better sensitivity to cases
    where those simple assumptions fail,
    such as in the isothermal assumption for Flare F2.
\autoref{fig:stix-decomp-vs-traditional} shows how uncertain the traditional spectroscopy
    fit is at fitting the low-energy portion of the thick target model.
This will influence the possible values of cutoff energy and electron flux,
    and $E_\text{nth} \propto \varphi_e E_C$\textemdash
    note, this convention is distinct from \cite{holman-acceleration-review},
    as their normalization cancels out a factor of the cutoff energy.
Furthermore,
    the nonthermal spectral index may be influenced by the low-error bins of counts in the thermal/nonthermal
    transition regions,
    even though the collection of higher-energy count bins should dominate the cold target fitting.
This has the effect of softening the photon spectrum,
    relative to the decomposition technique;
    this is reflected in the $\chi^2$ intervals annotated in \autoref{fig:stix-decomp-vs-traditional}
        and \autoref{fig:rhessi-decomp-vs-traditional}.
The spectrum may appear softer because traditional spectroscopy performs a kind of averaging
    over the analysis interval,
    while the time-decomposition technique extracts the behavior of impulsive acceleration events more accurately.

Furthermore,
    it would be difficult to
    apply the ``thick, warm'' target model
    \citep{kontar-thick-warm, emslie-thick-warm} to analyze time-decomposed data.
The energy and pitch angle diffusion in the thick+warm target model
    is perhaps more physically insightful than the traditional cold target assumption,
    and goes against the ``thermal'' vs ``nonthermal'' philosophy employed in the decomposition technique.
However,
    one may still compare physical parameters obtained from the thick/warm target model
    with those from the temporal decomposition technique.

\subsection{Cross-instrument analysis}

Because the temporal decomposition technique uses only timing information for the mathematical operations,
    we may combine information from different instruments\textemdash
    and even vastly different wavelengths\textemdash
    using this technique.
The temporal decomposition technique
    nondimensionalizes data during the fitting process,
    as noted immediately after \autoref{eq:tedec-norm}.
For the flares analyzed here,
    one could construct a Fermi/GBM spectrogram from time-tagged event data with very fine time bins
    and combine it with STIX or RHESSI data to perform a dual-instrument simultaneous fit.

Traditionally,
    simultaneous fitting is achieved by maximizing a joint objective function.
In the simultaneous temporal-decomposition case,
    simultaneous fitting would be achieved by first decomposing spectra in time
    and then maximizing one or more independent objective functions.
Various wavelengths such as microwave and radio data could be combined
    to form temporal pseudobases;
    for example,
    type III radio burst
    data from PSP/FIELDS \citep{fields-instr} could be used as
    part of a nonthermal basis,
    so long as the bursts are associated with prompt electron emission \citep{krucker-electrons}.
Microwave intensity curves from the Expanded Owens Valley Solar Array (EOVSA)
    could also be used as nonthermal bases.
Light curves extracted from SDO/AIA images could be used as a pseudobasis for
    the slowly-varying thermal component of the flare.
Such analyses would permit computations of the energy content of escaping
    electron beams associated with type III radio bursts,
    or investigating the properties of sunward vs. anti-sunward particle acceleration,
    as the anti-sunward accelerated particles lead to type III bursts.
The gyrosynchrotron emission measured by EOVSA is generally associated with higher-energy
    particles than those which produce bremsstrahlung X-rays,
    so if EOVSA intensity profiles can be used as nonthermal pseudo bases,
    this would support the hypothesis that lower- and higher-energy particles
    are accelerated in the same way,
    which could provide some clarity for traditional joint X-ray and microwave
    fitting as in e.g. \cite{bin-joint-xray-microwave}.

\section{Why this works: the multifractal formalism}
\label{sec:multifractal}
The temporal decomposition works because of differences in temporal behaviors
    between thermal plasma and accelerated electrons.
Here,
    we quantify the differences using the multifractal formalism,
    which is briefly described here,
    and defined more formally in \autoref{sec:appendix-multifrac}.
We use the multifractal formalism to compute a distribution of time series sharpness,
    and use the sharpness distribution to discern different physical emission mechanisms,
    taking heavy inspiration from \cite{mcateer-2007} who performed similar analysis.
Plots of function sharpness versus energy are constructed and explored
    intuitively in the context of thermal and nonthermal emission in solar flares.

Fractal systems are commonly found in nature and 
    can often be described by a single fractal dimension $D$
    which describes how much space an object fills\textemdash
    or how rough an object is.
A line is one dimension;
    self-similar shapes
    like the Sierpi\'{n}ski triangle
    fill more than one dimension,
    but not quite two dimensions.
In contrast to these simpler systems,
    solar flare X-ray time series, solar wind turbulent dynamics, DNA sequences,
    and the stock market
    \citep{mcateer-2007, solar-wind-mfdfa, multifractal-food-dna, mfdfa-stock-market}
    all exhibit behavior at multiple fractional dimension scales,
    and are therefore dubbed ``multifractal'' systems.
We may quantify the presence of each fractal scale via a multifractal spectrum.

Intuitively,
    a multifractal spectrum describes the distribution of function sharpnesses
    across a given time series.
The sharpness is quantified by a local scaling exponent
    called the \holder exponent $h$.

\begin{figure}
    \centering
        \includegraphics[width=\textwidth]{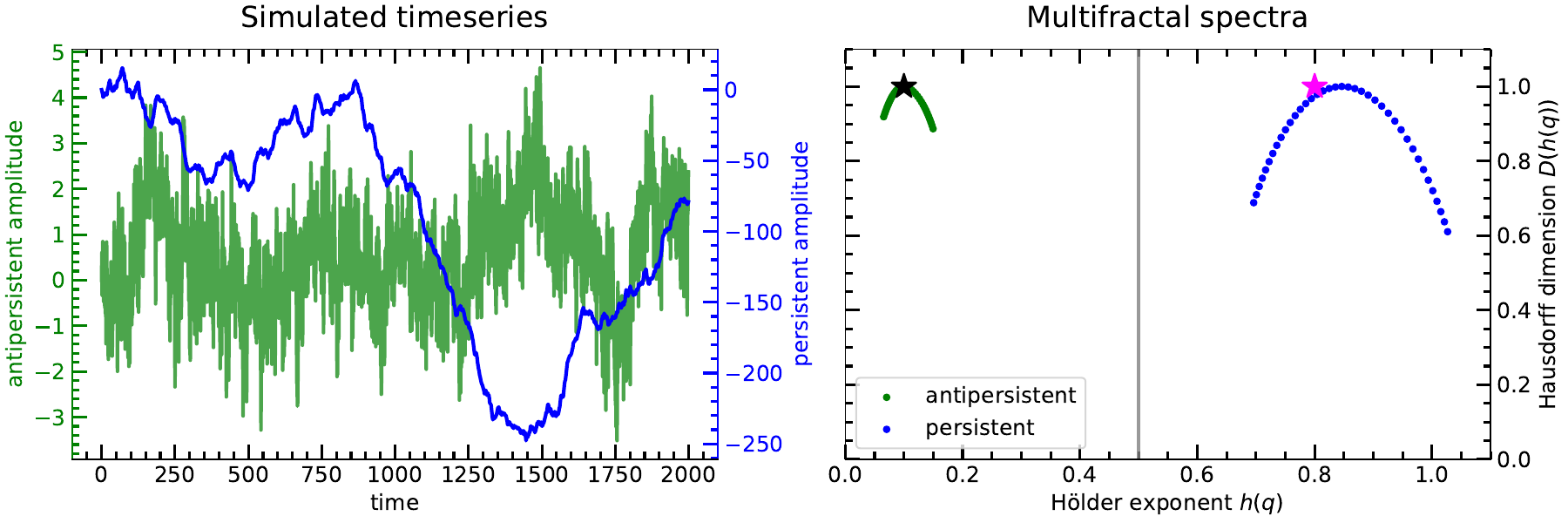}
    \caption{
        The MFDFA technique applied to two \textbf{simulated} time series.
        The time and amplitude values are in arbitrary units.
        The green data is ``antipersistent'',
            akin to Poisson noise present in X-ray detectors,
            and is defined as fractional Brownian motion (fBm) with a Hurst exponent $H = 0.1$
            \citep{mandelbrot-fbm}.
        The blue data is a ``persistent'' time series,
            which is more akin to thermal X-ray emission,
            defined as fBm with $H = 0.8$.
        The stars on the righthand plot indicate the ``ground truth'' time series Hurst exponents.
        An intuitive way to interpret the righthand plot is as a probability
            density of different fractal behavior across the time series.
        The lefthand time series are snapshots of infinite processes;
            in the infinite-time limit,
            the multifractal spectra on the right would converge to the stars.
        The gray line indicates the persistence/antipersistence boundary $h = 0.5$.
    }
    \label{fig:ex-multifrac}
\end{figure}

Consider the two time series (left)
    and resulting multifractal spectra (right) in \autoref{fig:ex-multifrac}.
The green time series stays closer to the $x$ axis and
    varies very quickly across time,
    like white noise;
    this is called antipersistent behavior and can be identified as a peak in the
    multifractal spectrum at $h < 0.5$.
The blue time series is more smoothly varying and strays further from the mean;
    this is called persistent behavior and corresponds to $h > 0.5$.
Prior work on this topic in the solar context has shown that
    thermal emission is generally more persistent,
    and nonthermal emission is more antipersistent \citep{mcateer-2007, mcateer-2013}.
Thermal emission grows and decays slowly compared to the measurement cadence
    as plasma heats and cools,
    while nonthermal emission fluctuates quickly on
    timescales of electron acceleration and propagation.
Photon shot noise in X-ray detectors is maximally antipersistent because it
    exhibits no correlation across time ($h \ll 0.5$).

To compute the multifractal spectra,
    we employ a technique known as multifractal detrended fluctuation analysis
    (MFDFA; \cite{mfdfa-og}).
Prior solar flare studies have employed a wavelet-based technique
    (e.g. \cite{mcateer-2007, mcateer-2013}) to quantify the multifractal spectra,
    but the MFDFA technique has been shown to be faster and more robust in practice \citep{cwtmm-mfdfa-compare}.
Additionally,
    MFDFA has solar physics heritage,
    as it was applied to analyze the fractal properties of solar wind turbulence \citep{solar-wind-mfdfa}.

Here,
    we briefly describe the steps to perform the MFDFA technique.
A less general form of MFDFA is known as detrended fluctuation analysis (DFA),
    which is used to quantify the Hurst exponent $H = 2 - d$\textemdash
    a monofractal scaling exponent,
    and $1 < d < 2$ a fractal dimension \citep{peng-dfa}.
To perform DFA,
    the time series is split into $N$ segments indexed by $s$ of length $n$.
The segments are fit with lines and the variance $\mathrm{Var}(n, s)$
    from the each is computed for each segment $s$.
The variance is normalized to define a fluctuation function $F(n, s) \sim \mathrm{Var}(n, s)^{1/2}$.
The fluctuation function (a measure of local complexity)
    depends on the length of the data segments $n$ (a measure of scale)
    and scales like $n^H$ where $H$ is the Hurst exponent.

DFA may be generalized to the multifractal sense by exploring different moments $q$
    of the fluctuation function, i.e.
    $F_\text{mf}(n, s) \sim \mathrm{Var}(n, s)^{1/q}$.
The moment $q$ weights short- or long-scale variations differently depending on its value,
    with $q = 2$ corresponding to the Hurst exponent calculation.
The multifractal fluctuation function $F_\text{mf}(n, s)$ scales as $n^{h(q)}$ with $h$ the \holder exponent \citep{mfdfa-og}.
So, from $F_\text{mf}(n, s)$, we may compute the multifractal spectrum.
We use the open-source \texttt{fathon} Python package to evaluate the multifractal spectra
    \citep{fathon-joss},
    which implements MFDFA as well as other fractal analysis techniques.

We apply the MFDFA technique to X-ray light curves of Flares F1 and F2
    as a function of energy.
Instead of plotting full multifractal spectra for each light curve,
    we plot only the dominant fractal behavior,
    which for \autoref{fig:ex-multifrac} would correspond to
        points plotted close to the magenta and black stars.

\begin{figure}
    \centering
        \includegraphics[width=\textwidth]{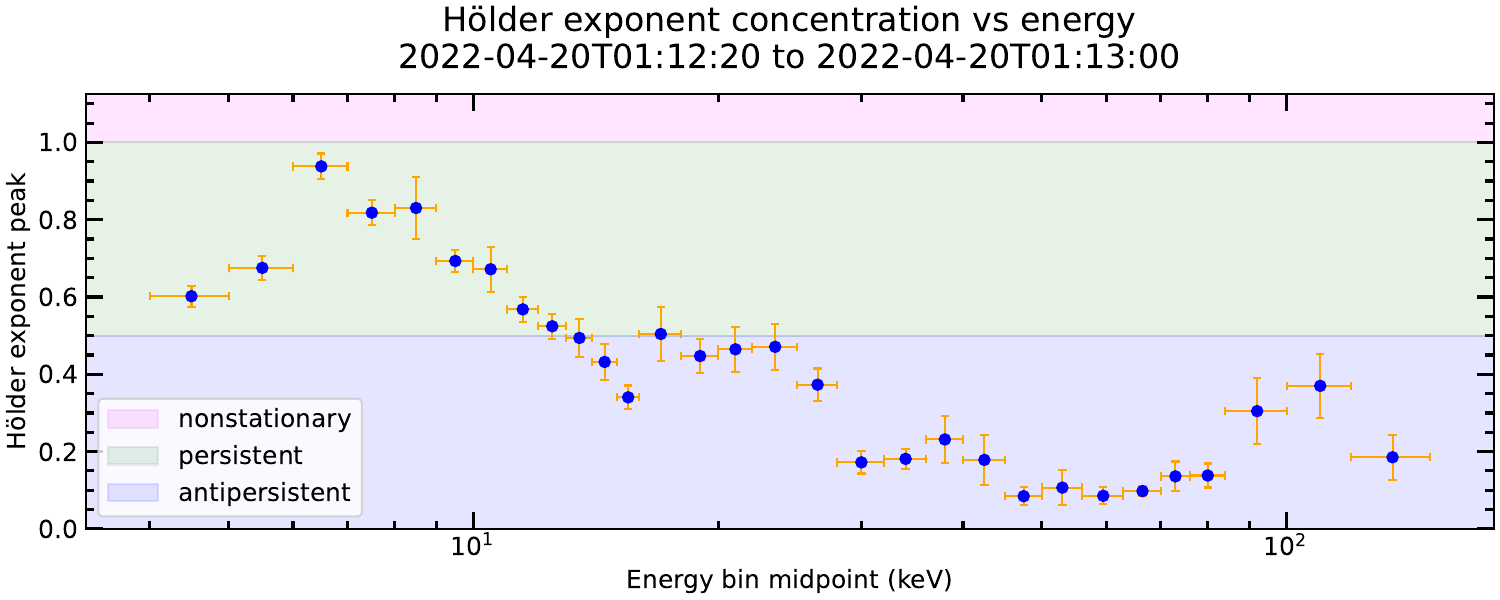}
    \caption{
        Multifractal spectrum peak and spread as a function of energy for Flare F1.
        We only focus on the most prevalent behavior of each timeseries,
            which corresponds to multifractal moments $q = 0$ ($y$ value) and $q = \pm 0.2$ ($y$ error);
            see \citep{mfdfa-og} for information on the interpretation of the moments.
        The $x$ errorbars correspond to energy bin widths.
        Counts data that are background-dominated (i.e. mostly Poisson noise)
            are very antipersistent and stay entirely in the blue region
            ($E \ge 28$ keV).
    }
    \label{fig:stix-mf-justify}
\end{figure}

For flare F1, the behavior is plotted in \autoref{fig:stix-mf-justify}.
The colored regions correspond to three different timeseries behaviors:
    antipersistent,
    persistent,
    and nonstationary (i.e. non-mean-reverting).
The energy bands dominated by thermal emission are mainly persistent verging on nonstationary;
    those by nonthermal are mainly antipersistent verging on persistent;
    and, the thermal/nonthermal energy cross-over is characterized by a transition from
    persistent to antipersistent.
From this summary plot,
    we may infer that a good choice of pseudobasis vectors for this flare
    would be any persistent low-energy band for the thermal pseudobasis,
    and any energy band which has approximately the same dominant
    \holder exponent around 20 keV for the nonthermal basis.
Notably, we did not include any spectroscopic or energy information in this analysis.
The fractal information is purely a function of light curve shape.

\begin{figure}
    \centering
        \includegraphics[width=\textwidth]{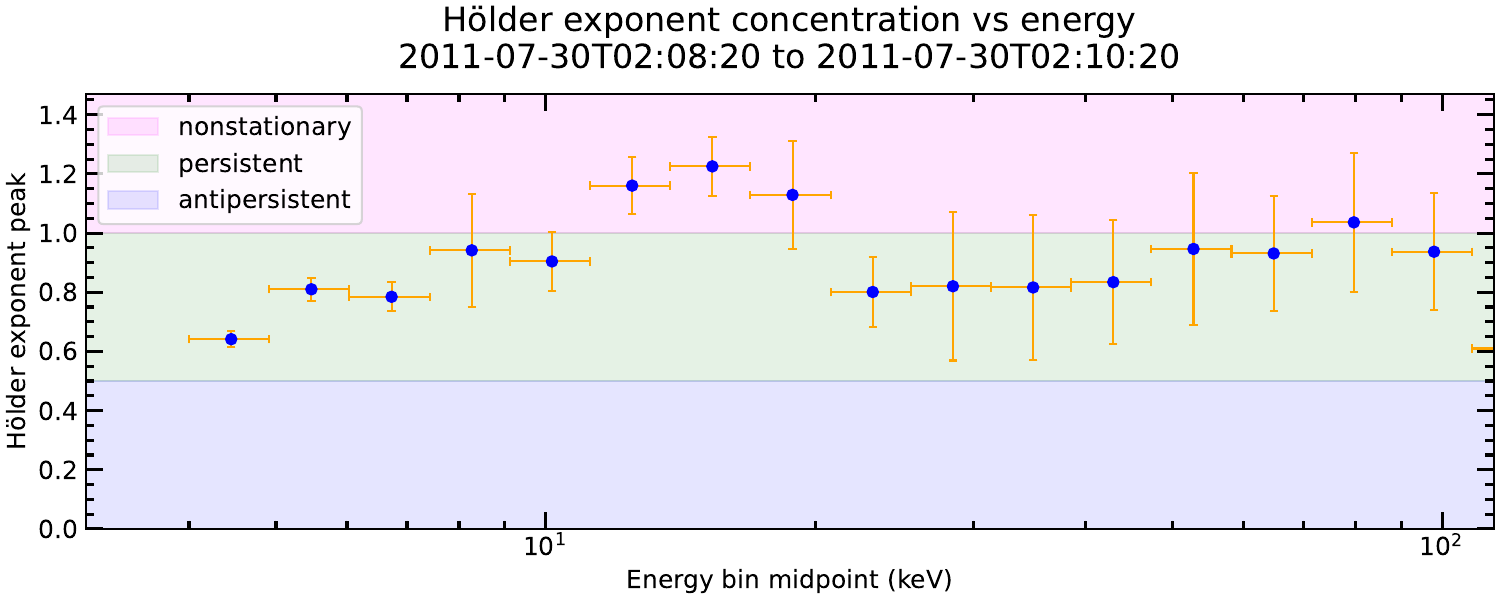}
    \caption{
        Multifractal summary for Flare F2, produced in the same manner as \autoref{fig:stix-mf-justify}.
        We see similar behavior as in the case of the STIX flare,
            but the nonthermal emission remains persistent even at higher energies,
            and the $y$ error bars are larger.
        This is because there are fewer time bins across a similar analysis interval,
            and the higher energies are not dominated by shot noise
            from a radioactive calibration source.
    }
    \label{fig:rhessi-mf-justify}
\end{figure}

A more precise interpretation of the multifractal summary plots is,
    ``clumps of energy bins in the fractal summary plots
      which all have similar \holder exponents are likely
      full of counts from the same physical emission mechanism.''
\autoref{fig:rhessi-mf-justify} shows a similar summary plot for Flare F2,
    which was observed by RHESSI.
Here again we see that for low energies,
    the time series are persistent to nonstationary.
For higher energies the emission is persistent,
    but the \holder exponent $h$ error bars are larger here,
    in part due to the coarser time binning (fewer data points).
Again we see a transition region around ~20 keV which nicely correlates
    with the thermal/nonthermal transition region inferred from spectroscopy.
There appear to be two distinct \holder exponent clumpings at low vs. high nonthermal energies:
    one from 25-50 keV,
    and one from 50-100 keV.
Perhaps these clumps correspond to two different electron populations.
Again we may choose our nonthermal pseudobases based on these summary plots.

Interestingly,
    no statistically significant periodic pulsations were identified in either flare during
    the spectroscopy intervals.
Quasi-periodic pulsations were investigated using the AFINO package \citep{afino-first, afino-second}.
A sympathetic flare was observed after Flare F1 at around 01:40 UTC,
    and this flare did exhibit significant oscillations with period $\sim$27 s.
Therefore,
    periodic behavior is not required to apply this decomposition technique,
    and is not necessary to make use of the multifractal spectra.

The dominant fractal behavior of X-ray light curves may be used
    to select and justify the choice of time pseudobases,
    and give an early indication of whether or not the method will work well.
These techniques may be applied to any spectrogram-like data,
    whether it be X-ray,
    radio,
    particle,
    or otherwise.
You don't need to use the multifractal formalism to
    get started with the temporal decomposition technique,
    but it is a robust way to check if you have picked good pseudobases.

\section{Conclusion}
We have developed and demonstrated a new technique for performing spectroscopy on solar flare X-ray data.
Here are three main takeaways from this work:
\begin{enumerate}
    \item Timing information can be used to aid in spectroscopy,
          lifting energy degeneracies which typically lead to large uncertainties,
          as it permits fitting thermal/nonthermal spectra independently.
    \item Time-informed spectroscopy yields (sometimes orders of magnitude) more precise
          results than traditional spectroscopy,
          which impacts calculations such as the total energy contained in flare electrons.
    \item The multifractal formalism gives a robust definition of ``time series shape,''
          and may be used to select and/or verify pseudobasis light curves.
\end{enumerate}

In future work,
    we plan to apply this technique to other wavelengths and data products,
    such as ground- and space-based radio observations,
    and space-based particle observations.
We encourage readers to apply this technique to other subfields of physics,
    and entirely different domains.

\begin{acknowledgements}
We would like to acknowledge help from James T. McAteer with debugging
    an earlier version of the multifractal formalism code,
    and for offering his own implementation of the wavelet transform modulus maxima method.
We thank Andy Inglis for his help searching for quasiperiodic pulsations in our selected flares,
    and we thank him and Trevor Knuth for conversations pertaining to this work.

We acknowledge NSF CAREER grant AGS 1752268 and NASA grant 80NSSC20K1318
    for funding this work.
\end{acknowledgements}

\bibliographystyle{aasjournalv7}
\bibliography{main}{}

\appendix
\section{Details: the multifractal formalism}
\label{sec:appendix-multifrac}

Multifractal systems can be described by a distribution of Hausdorff dimensions
    plotted as a function of \holder exponent;
    this is the multifractal spectrum.
Intuitively,
    the multifractal spectrum quantifies how local function sharpness (\holder exponent)
    fills the space (Hausdorff dimension) of the timeseries.
The multifractal spectrum can be interpreted as a probability density
    $p(h(q))$ of \holder exponents.
Here we define the \holder exponent and give a pictoral
    representation of how \holder exponents envelop the local behavior of time series.
We then define the Hausdorff dimension for a generic metric space and relate it
    to the multifractal analysis.

The \holder exponent is an example of a singularity exponent.
The singularity exponent describes the local function sharpness and
    is defined as in \autoref{eq:singularity},
    where $t$ is the independent variable,
    $f(t)$ is the function under analysis,
    $\tau$ is some fixed time,
    and $p(t)$ is the time-dependent singularity exponent enveloping the function
    about $t = \tau$ \citep{mcateer-2007, feder-fractals}.
\begin{equation}
    f(t - \tau) \sim \tau^{p(t)}
    \label{eq:singularity}
\end{equation}

\begin{figure}
    \centering
        \includegraphics[width=0.8\textwidth]{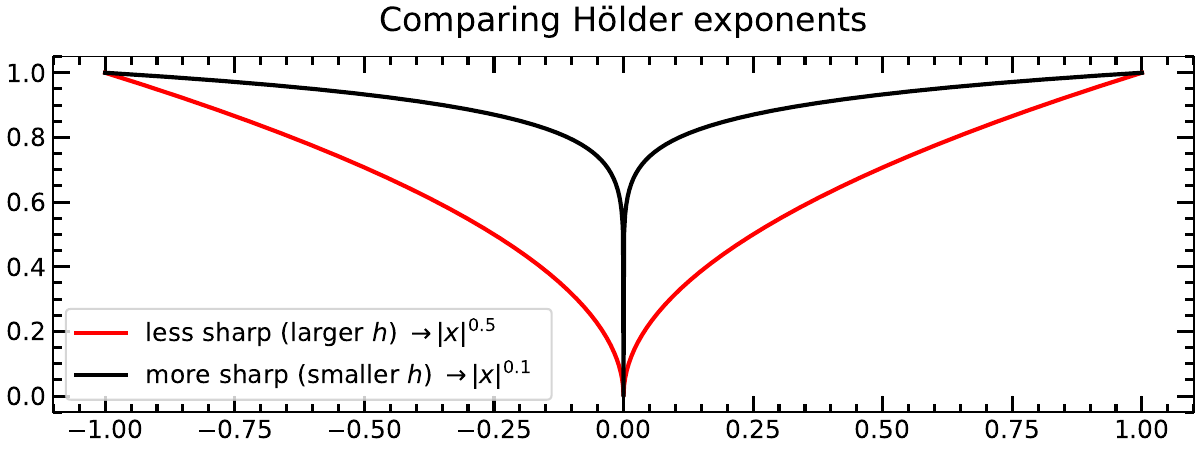}
    \caption{
        Comparison of local \holder exponent envelopes about $x = 0$.
        Sharper features are described by smaller exponents.
    }
    \label{fig:cmp_hoelder}
\end{figure}

The \holder exponent is a singularity exponent which has useful properties.
First we define the set $C^\alpha(\tau)$ as in \autoref{eq:calpha}
    for a polynomial $P_m$ of degree $m$, $m < \lfloor\alpha\rfloor$,
    at some time $t = \tau$.
\begin{equation}
    \left| f(t) - P_m(t - \tau) \right|
    \le
    C\left| t - \tau \right|^\alpha
    \label{eq:calpha}
\end{equation}
The timeseries $f \in C^\alpha(\tau)$ when \autoref{eq:calpha} is satisfied;
    that is,
    when the local time series behavior scales like a power law on top of
    an $m$\textsuperscript{th} order polynomial.
Then, the \holder exponent $h(f, \tau)$ is defined as in equation \autoref{eq:hoelder}.
\begin{equation}
    h(f, \tau) = \mathrm{sup} \left\{ \alpha > 0 \mid f \in C^\alpha(\tau) \right\}
    \label{eq:hoelder}
\end{equation}
The \holder exponent is the supremum (least upper bound) $\alpha$ that describes the
    local scaling of the timeseries $f(t \approx \tau)$ \citep{mcateer-2007}.
\textbf{Because the local scaling is a power law,
    smaller $h(\tau)$ describe sharper
    features than larger $h(\tau)$ as illustrated in \autoref{fig:cmp_hoelder}.}
For fractal time series,
    it is usually the case that $0 < \alpha < 1$,
    but not always.

The other dimension in the multifractal spectrum is on its $y$ axis,
    the Hausdorff dimension.
Informally,
    the Hausdorff dimension describes how much space a measurable object fills.
In the multifractal context,
    it can be interpreted as a probability distribution of \holder exponents.

We may define the Hausdorff dimension on any metric space,
    in particular that of \holder exponents on a time series.
Let $X$ be a metric space\textemdash
    a set of points with a metric function;
    for example,
    the set of real numbers $\mathbb{R}$ is a metric space with $x, y \in \mathbb{R}$
    and metric function $g(x, y) = |x - y|$.
Let $S \subset X$ and the diameter $\mathrm{diam}(S) = \sup\{g(x, y) \mid x, y \in S\}$.

Define the incomplete Hausdorff measure for the set $S$ as in \autoref{eq:hausdorff-meas-incomp}.
$I$ is some (potentially infinite) indexing set,
    $U_i$ is a simple shape which covers part of the set $S$,
    $\delta$ is the maximum allowed diameter,
    and $d$ is a power which weights the summation (the ``moment'' of the sum).
The equation sums up the $d$\textsuperscript{th} powers of the diameters of simple shapes\textemdash
    balls, boxes, or otherwise\textemdash
    covering the metric space $S$.
The maximum diameter $\delta$ may be made arbitrarily small.
\begin{equation}
    H_\delta^d(S) = \inf
        \left\{
            \sum_{i \in I} \left(\mathrm{diam}(U_i)\right)^d
            :
            \bigcup_{i \in I} U_i \supseteq S,\ \mathrm{diam}(U_i) < \delta
        \right\}
    \label{eq:hausdorff-meas-incomp}
\end{equation}
Now define the (complete) Hausdorff measure as in \autoref{eq:hausdorff-meas}.
\begin{align}
    H^d(S) &= \sup_{\delta > 0} H_\delta^d(S) \nonumber \\
           &= \min_{\delta > 0} H_\delta^d(S) \nonumber \\
           &\equiv \lim_{\delta \to 0} H_\delta^d(S) \label{eq:hausdorff-meas}
\end{align}
The Hausdorff measure is $H^d(S)$.
There exists some ``transitional'' $d$ for which the Hausdorff measure goes from
    $H^d(S) = \infty$ to $H^d(S) = 0$
    which may be found by evaluating the limit $\delta \to 0$.

From the Hausdorff measure define the Hausdorff dimension,
    $D(S)$ as in \autoref{eq:hausdorff-dim}.
\begin{equation}
    D(S) = \inf\left\{d \ge 0 : H^d(S) = 0\right\}
    \label{eq:hausdorff-dim}
\end{equation}
In the MFDFA case we approximate the Hausdorff dimension as $D(h(q))$ as in \cite{fathon-joss},
    where the metric function on the set $h(q)$ is some distance between \holder exponents
    dependent upon the local function sharpnesses
    and fluctuation function $F_\text{mf}(n, s)$ as described in \autoref{sec:multifractal}.
\cite{mfdfa-og}, section 2.2, describes the mathematical equivalence of the fluctuation function to
    the way folks traditionally compute the multifractal spectrum,
    which is via the so-called box-counting method.
There, the balls $U_i$ are replaced by boxes and a statistical mechanics approach is applied to
    define the fractal scaling (the $p(t)$ from \autoref{eq:singularity})
    as a function of the number of boxes it takes
    to cover parts of the time series at different box sizes.

\section{Corner plots: traditional vs. decomposed spectroscopy}
\label{sec:appendix-corners}
Corner plots offer a fast way to visualize parameter covariances and marginal posterior distributions.
All corner plots produced in this work were made using \texttt{corner.py} \citep{corner-py-joss}.

95\% posterior intervals are annotated on all of the marginal distributions in the corner plots
    as dashed lines.
The titles have the (2.5\%, 50\%, 97.5\%) quantiles annodated in
    $P^{+ \delta P_1}_{- \delta P_2}$ fashion,
    with $P$ a parameter value and $\delta P_i$ the one-sided uncertainty.
The values annotated are the same as the 95\% intervals given in the text,
    just written in a different way.
The median values of the posteriors are annotated in the titles 
    in order to be more congruous with the histograms.
The best-fit values are annotated as the red crosses in each plot.

\begin{figure}
    \centering
        \includegraphics[width=\textwidth]{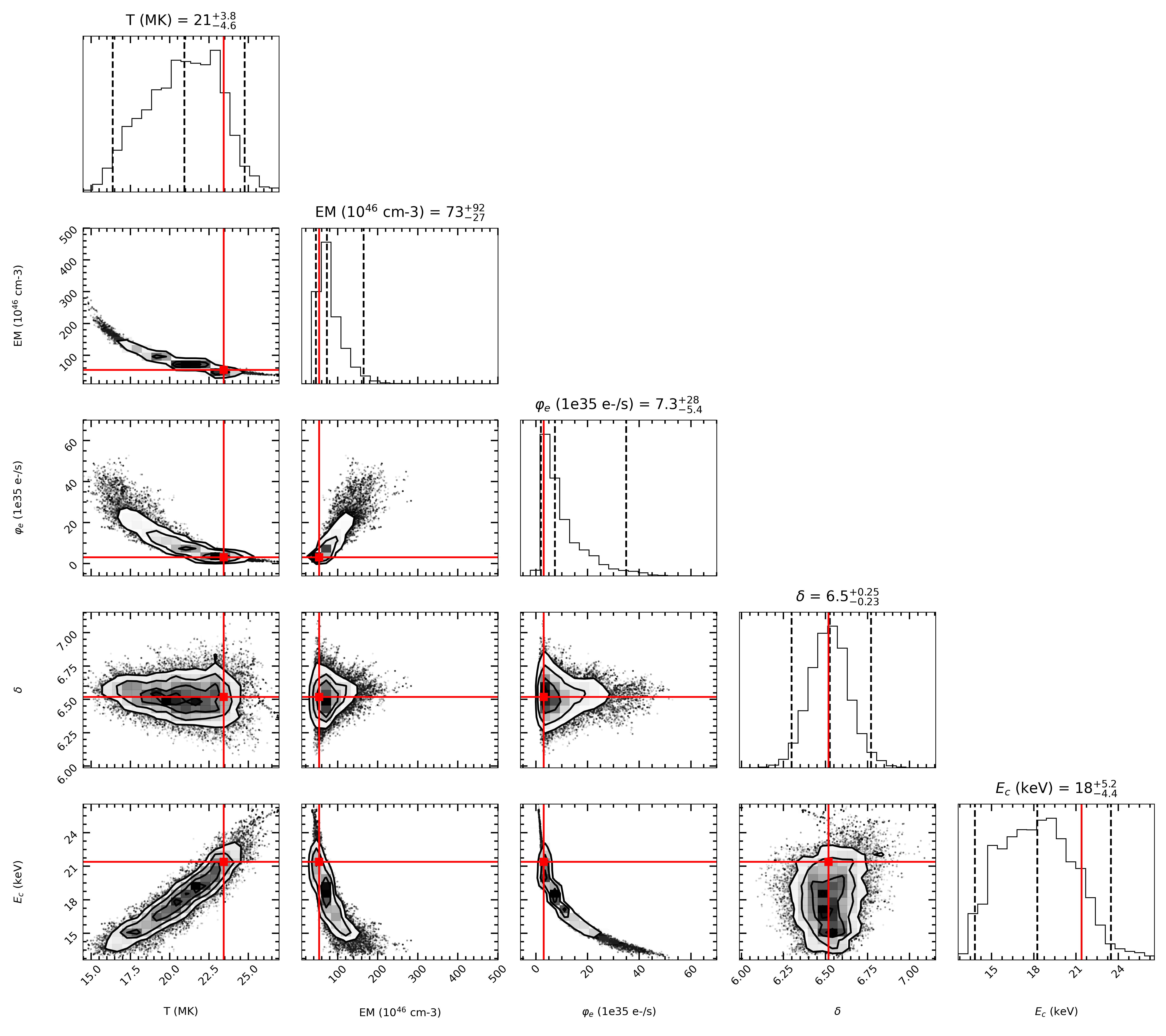}
    \caption{
        Corner plots for parameters fit using traditional spectroscopy
        on Flare F1 across its analysis interval.
        The red crosshairs mark the best fit parameters
            found via Levenberg-Marquadt minimization.
    }
    \label{fig:corner-stix-full}
\end{figure}

\begin{figure}
    \centering
        \includegraphics[width=\textwidth]{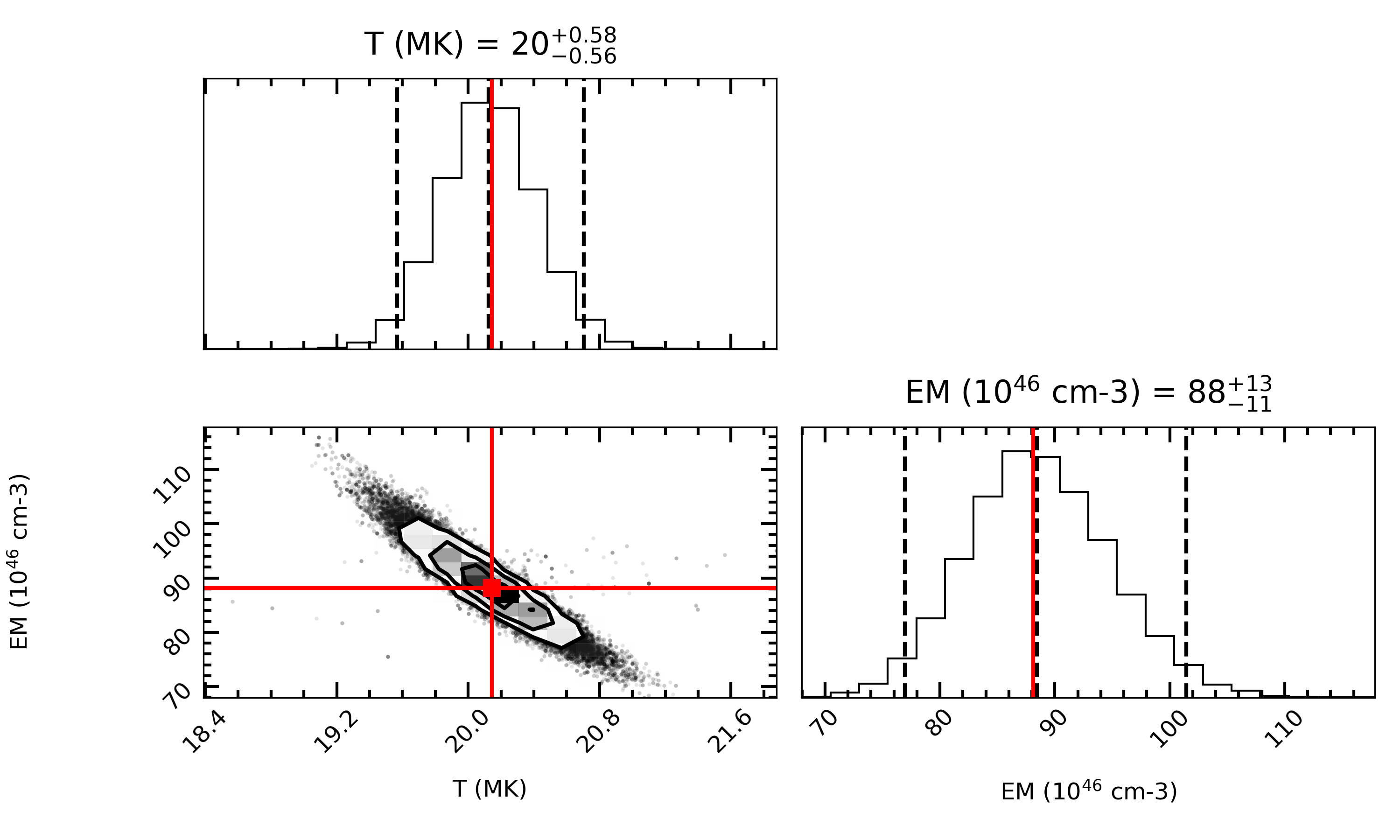}
    \caption{
        Corner plots for thermal parameters fit using time-decomposed spectroscopy
        on Flare F1 across its analysis interval.
        The red crosshairs mark the best fit parameters
            found via Levenberg-Marquadt minimization.
    }
    \label{fig:corner-stix-thermal}
\end{figure}

\begin{figure}
    \centering
        \includegraphics[width=\textwidth]{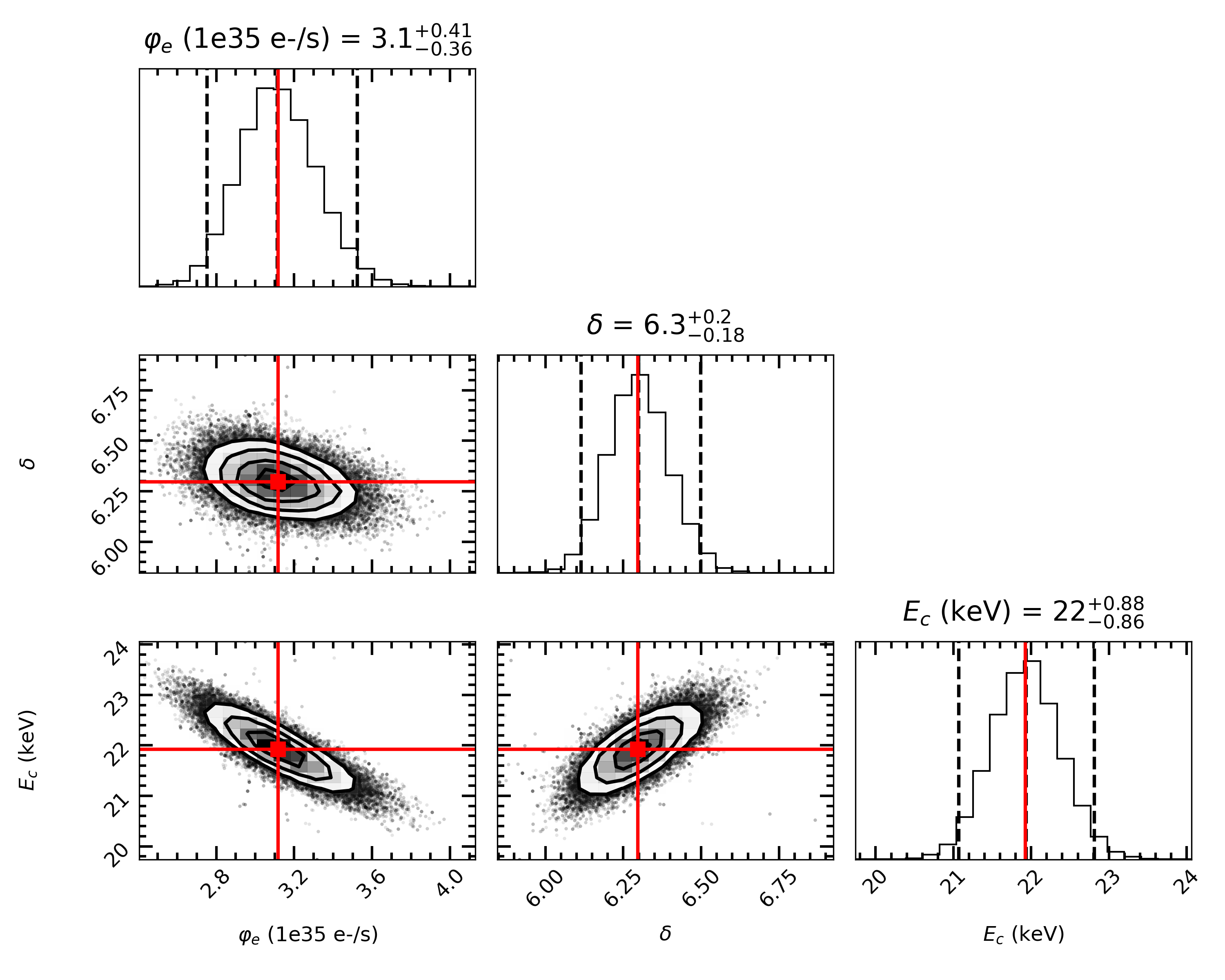}
    \caption{
        Corner plots for nonthermal parameters fit using time-decomposed spectroscopy
        on Flare F1 across its analysis interval.
        The red crosshairs mark the best fit parameters
            found via Levenberg-Marquadt minimization.
    }
    \label{fig:corner-stix-nonthermal}
\end{figure}

\begin{figure}
    \centering
        \includegraphics[width=\textwidth]{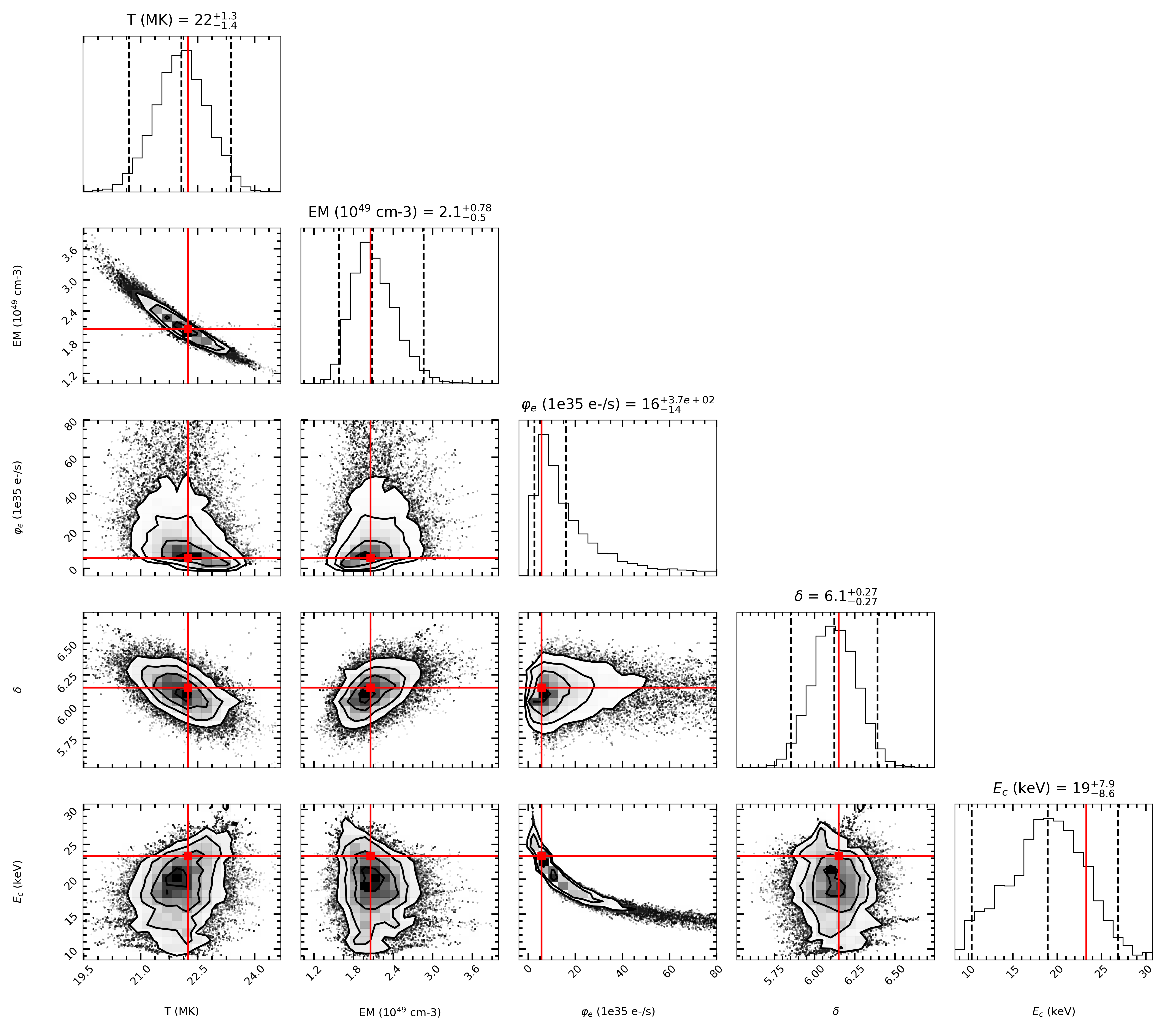}
    \caption{
        Corner plots for parameters fit using traditional spectroscopy
        on Flare F2 across its analysis interval.
        The red crosshairs mark the best fit parameters
            found via Levenberg-Marquadt minimization.
    }
    \label{fig:corner-rhessi-full}
\end{figure}

\begin{figure}
    \centering
        \includegraphics[width=\textwidth]{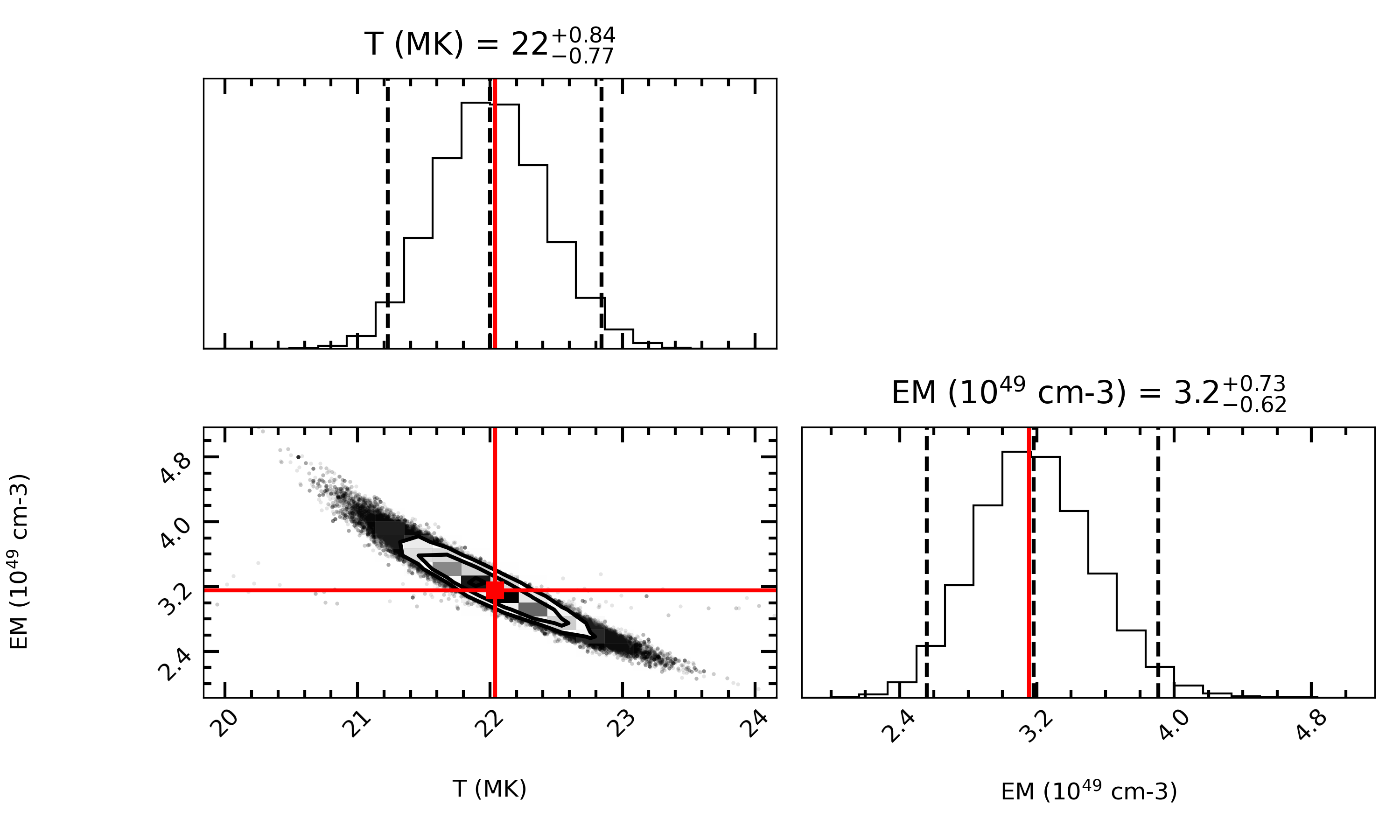}
    \caption{
        Corner plots for thermal parameters fit using time-decomposed spectroscopy
        on Flare F2 across its analysis interval.
        The red crosshairs mark the best fit parameters
            found via Levenberg-Marquadt minimization.
    }
    \label{fig:corner-rhessi-thermal}
\end{figure}

\begin{figure}
    \centering
        \includegraphics[width=\textwidth]{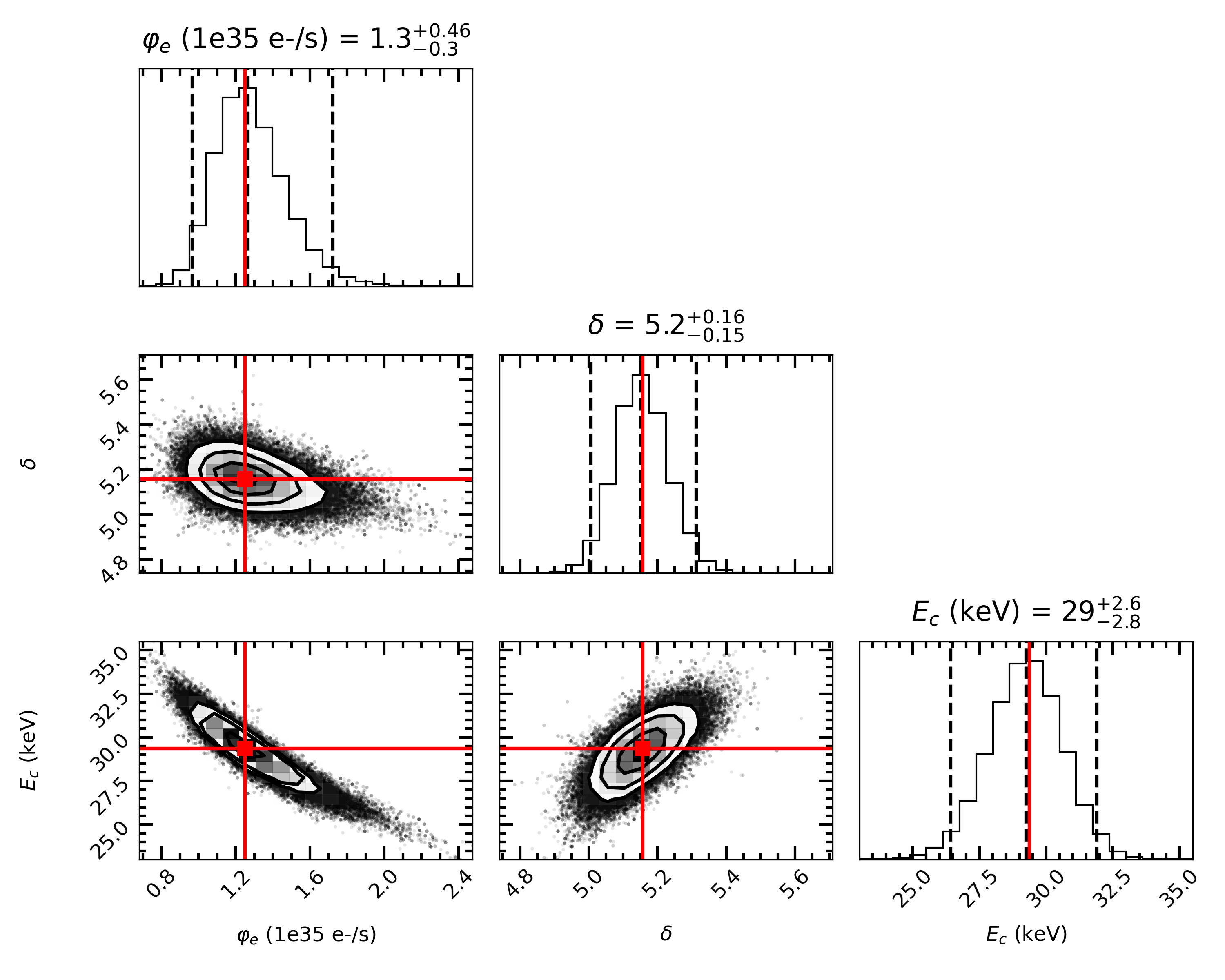}
    \caption{
        Corner plots for nonthermal parameters fit using time-decomposed spectroscopy
        on Flare F2 across its analysis interval.
        The red crosshairs mark the best fit parameters
            found via Levenberg-Marquadt minimization.
    }
    \label{fig:corner-rhessi-nonthermal}
\end{figure}

\end{document}